\documentclass[a4paper,onecolumn,accepted=2022-05-19]{quantumarticle}

\pdfoutput=1

\usepackage[utf8]{inputenc}
\usepackage[english]{babel}
\usepackage[a4paper, margin=1in]{geometry}
\usepackage{graphicx}
\usepackage{amsmath}
\usepackage{amsfonts}
\usepackage{pgfplots}
\pgfplotsset{compat=1.16}
\pgfplotsset{scaled x ticks=false}
\definecolor{cb-blue}{RGB}{0,119,187}
\definecolor{cb-cyan}{RGB}{51,187,238}
\definecolor{cb-teal}{RGB}{0,153,136}
\definecolor{cb-orange}{RGB}{238,119,51}
\definecolor{cb-red}{RGB}{204,51,17}
\definecolor{cb-magenta}{RGB}{238,51,119}
\definecolor{cb-grey}{RGB}{187,187,187}
\usepackage{physics}
\usepackage{bbold}
\usepackage{bm}
\usepackage{subcaption}
\usepackage{algorithm}
\usepackage{algpseudocode}

\usepackage[numbers,sort&compress]{natbib}
\usepackage{xcolor}
\usepackage{hyperref}
\hypersetup{
    colorlinks,
    linkcolor={red!50!black},
    citecolor={green!50!black},
    urlcolor={blue!50!black}
}

\title{Numerical Implementation of Just-In-Time \newline Decoding in Novel Lattice Slices Through the Three-Dimensional Surface Code} 
\author[1,2]{T. R. Scruby}
\author[2]{D. E. Browne}
\author[3]{P. Webster}
\author[4,5]{M. Vasmer}
\affil[1]{Okinawa Institute of Science and Technology, Okinawa, 904-0495, Japan}
\affil[2]{Dept. of Physics and Astronomy, University College London, London, WC1E 6BT, UK}
\affil[3]{Centre for Engineered Quantum Systems, School of Physics,\\ The University of Sydney, Sydney, NSW 2006, Australia}
\affil[4]{Perimeter Institute for Theoretical Physics, Waterloo, ON N2L 2Y5, Canada}
\affil[5]{Institute for Quantum Computing, University of Waterloo, Waterloo, ON N2L 3G1, Canada}

\begin{document}

\maketitle

\begin{abstract}
    We build on recent work by B. Brown (Sci.\ Adv.\ 6, eaay4929 (2020)) to develop and simulate an explicit recipe for a just-in-time decoding scheme in three 3D surface codes, which can be used to implement a transversal (non-Clifford) $\overline{CCZ}$ between three 2D surface codes in time linear in the code distance. We present a fully detailed set of bounded-height lattice slices through the 3D codes which retain the code distance and measurement-error detecting properties of the full 3D code and admit a dimension-jumping process which expands from/collapses to 2D surface codes supported on the boundaries of each slice. At each timestep of the procedure the slices agree on a common set of overlapping qubits on which $CCZ$ should be applied. We use these slices to simulate the performance of a simple JIT decoder against stochastic $X$ and measurement errors and find evidence for a threshold $p_c \sim 0.1\%$ in all three codes. We expect that this threshold could be improved by optimisation of the decoder. 
\end{abstract}

\section{Introduction and Overview}
\label{section:intro}
Quantum error correcting codes allow us to protect quantum systems against decoherence~\cite{shor1995,steane1996a} and are believed to be a necessary part of a scalable quantum computer~\cite{campbell2017b}. 
Topological codes~\cite{kitaev2003,bombin_topological_2007} are an important class of codes with geometrically local stablisers and high error tolerance~\cite{raussendorf2007,wang2011}. 
Topological codes that are scalable in two dimensions (2D) are of particular relevance, as many state-of-the-art quantum processors share this constraint~\cite{barends2014,Arute2019,jurcevic_demonstration_2021}.
An important question in any quantum error correction scheme is how to fault-tolerantly implement a universal gate set for a given code. 
No-go results have been proved that limit naturally fault-tolerant operators from being universal in topological codes~\cite{eastin2009,zeng2011,bravyi2013a,jochym-oconnor2018,webster_universal_2022,webster_fault-tolerant_2020}. 
Nonetheless, procedures exist for implementing a universal gate set in 2D topological codes. These include a recent proposal by Brown~\cite{brown_fault-tolerant_2020}, the implementation of which is the subject of this work.

A decoding algorithm is a classical process in a quantum error correction scheme which takes as input a set of stabiliser measurement outcomes (referred to as a syndrome) and returns a correction operator. A correction is successful if the product of the correction operator and the original error is equivalent to a stabiliser. Of particular interest are \textit{single-shot} decoding schemes where corrections can be inferred reliably from a single round of stabiliser measurements even in the presence of measurement errors~\cite{bombin_single-shot_2015,campbell_theory_2019,quintavalle_single-shot_2021}. It is believed that single-shot decoding is not possible for 2D topological codes~\cite{campbell_theory_2019} and so measurement errors must instead be counteracted by repeated rounds of stabiliser measurement, with fault-tolerance in a distance $d$ code requiring $O(d)$ repeats~\cite{fowler_surface_2012}. As noted by Bomb\'{i}n in~\cite{bombin_2d_2018} this need for repeated measurements results in a discrepancy between constant-time circuits in 3D topological codes and linear-time circuits in 2D topological codes: in principle these both have spacetime cost $O(d^3)$, but in practise the 2D code will also require $O(d)$ measurement rounds between each set of operations, resulting in a time cost of $O(d^2)$ and a spacetime cost of $O(d^4)$. 

To rectify this, Bomb\'{i}n proposed the concept of a just-in-time (JIT) decoder~\cite{bombin_2d_2018} which supplies, at each timestep of the computational procedure, a best-guess correction based not only on the present syndrome of the code but also on the entire syndrome history. By interpreting the syndrome of a (2+1)-dimensional code as the syndrome of the corresponding (3+0)-dimensional code such a decoder allows for a form of pseudo-single-shot decoding of 2D codes, where measurements are not repeated and mistakes in the correction due to measurement errors are compensated for at later timesteps once the measurement errors that caused them are identified. The price for using such a decoder is that our 2D codes must be replaced with very thin slices of 3D code in which we can detect (but not reliably correct) measurement errors. The thickness of these slices is independent of $d$ and so while they are not strictly 2D (in the sense that they cannot be embedded in a two-dimensional manifold) they still only require an architecture which is scalable in just two dimensions. In what follows, we will use the term ``layer'' to refer to a strictly 2D code and ``slice'' to refer to a bounded-thickness section of 3D code. We will also use bars over states and operators to refer to logical versions of these. 

In its original formulation, Bomb\'{i}n used the idea of JIT decoding to circumvent causal restrictions encountered when attempting to translate a (3+0)-dimensional measurement-based computing scheme to a (2+1)-dimensional one. This scheme was based on the 3D colour code~\cite{bombin_topological_2007, bombin_topological_2007-1,kubica2015a} and the ideas presented there were translated to the 3D surface code by Brown~\cite{brown_fault-tolerant_2020}, who used JIT decoding to prove a threshold for a linear-time $\overline{CCZ}$ between three 2D surface codes. When combined with the Clifford group (which is also implementable in the 2D surface code~\cite{fowler_surface_2012, brown_poking_2017}) this provides us with a way to obtain a universal gate set which is potentially more efficient than competing techniques such as magic state distillation~\cite{fowler_surface_2012,bravyi2005,Litinski2019magicstate}.  

In this work, we present a full implementation of Brown's procedure, which incorporates several new techniques to refine it. In each of the three codes, we construct a scalable slice which is compatible with the various requirements of the procedure (discussed below).  Having constructed these slices, we then simulate the performance of a simple JIT decoder in this setting and observe a threshold $p_c \sim 0.1\%$ in all three codes; see Fig.~\ref{fig:threshold_plots}. This is (to our knowledge) the first numerical demonstration of a threshold for JIT decoding. 

In what remains of this section we provide an overview of our implementation of Brown's procedure and then discuss each component in more detail in subsequent sections. 

The first such component is the 3D surface code (Section~\ref{section:3DSC}). Unlike its 2D counterpart, this code admits a transversal three-qubit non-Clifford gate (the $CCZ$ gate) between three overlapping copies of the code~\cite{kubica_unfolding_2015,vasmer_three-dimensional_2019}. The aim of Brown's procedure is to use the equivalence between the 3D surface code in (3+0) dimensions and the 2D surface code in (2+1) dimensions to implement a linear-time (in the code distance $d$) version of this gate between three overlapping 2D codes. 

To achieve this we require a division of the three overlapping 3D codes into $O(d)$ bounded height slices which satisfy various requirements, the most important ones being that each slice must itself be a valid code with distance $d$, and that at each timestep the three codes should agree on a common set of qubits on which we should apply $CCZ$ (Section~\ref{section:codes}). We also require a method of moving from one slice to the next, which cannot simply be ``waiting'' (despite the fact that we are using time as a dimension) because the overlap of the three 2D codes must be different at each timestep so they must move relative to each other. This is achieved by \textit{dimension jumping}~\cite{bombin_dimensional_2016} between the slice and the 2D layers on its top and bottom. $Z$ errors which arise during the procedure are also dealt with as part of this process (Section~\ref{section:dimension_jumping}). We must also ensure that all of these operations commute with the logical action of the $CCZ$ so that the entire procedure has the intended effect (Section~\ref{section:CCZ}).

If the above can be done successfully then we will have obtained a procedure which, in the absence of errors, implements $\overline{CCZ}$ between three 2D surface codes in linear-time. To make the procedure fault-tolerant in the presence of $X$ and measurement errors we must use a just-in-time decoder (Section~\ref{section:decoder}). A single timestep of the full, fault-tolerant procedure then occurs as follows:

\begin{itemize}
    \item Begin with three overlapping 2D codes/layers
    \item Expand to three overlapping slices of 3D code
    \item Apply JIT decoding operations to the three slices
    \item Apply $CCZ$ gates between the overlapping qubits
    \item Collapse back to three 2D layers 
\end{itemize}

The layers we collapse to are on the opposite side of the slice from those we started in. It is possible to implement the entire procedure on an architecture which is only one slice thick by redefining our time direction at the start of each timestep.

\section{The 3D Surface Code}
\label{section:3DSC}
\begin{figure}
    \begin{subfigure}{.5\textwidth}
        \includegraphics[width=.9\textwidth]{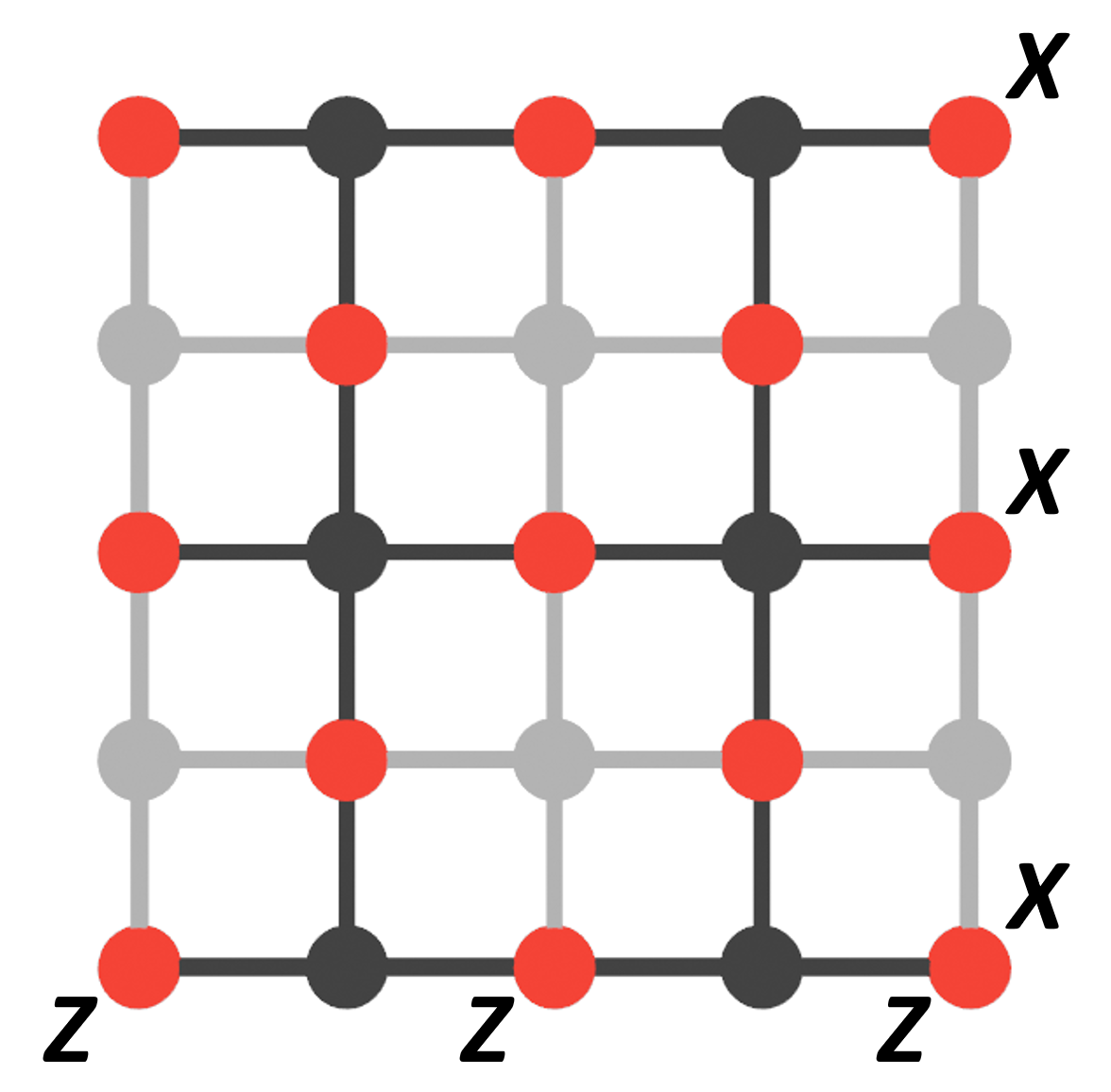}
        \subcaption{}
    \end{subfigure}
    ~~
    \begin{subfigure}{.5\textwidth}
        \includegraphics[width=.9\textwidth]{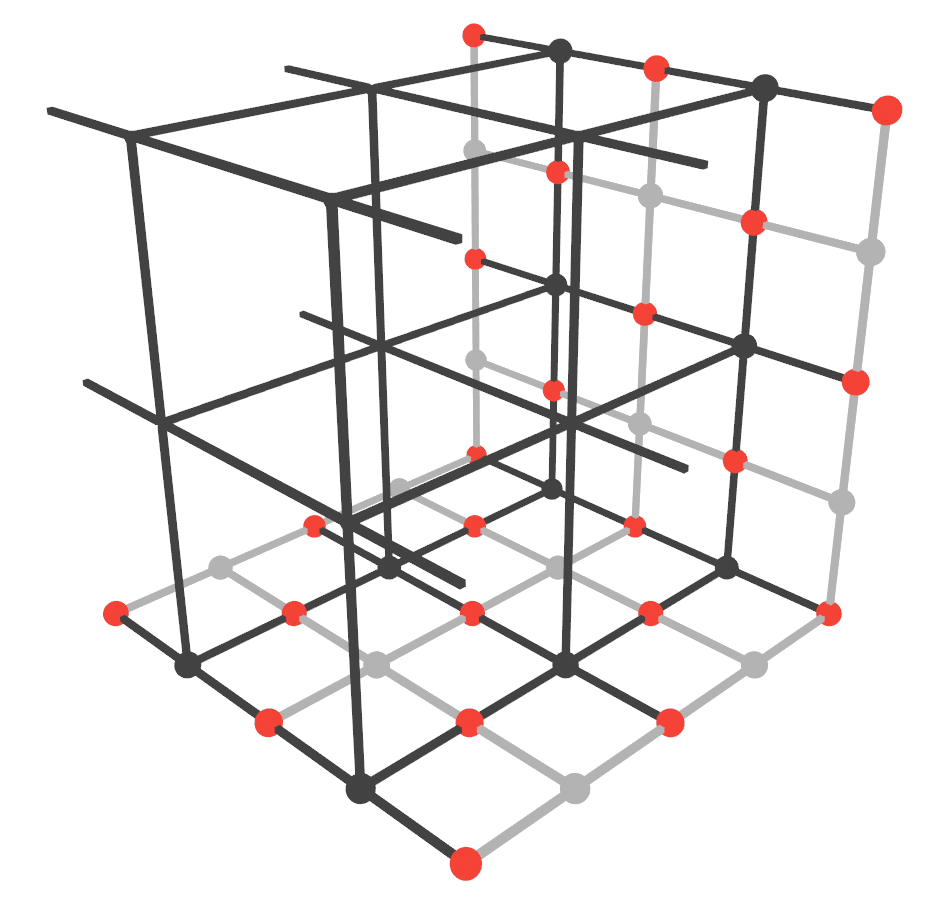}
        \subcaption{}
    \end{subfigure}
    \caption{(a) A 2D surface code. $X$ stabilisers are on vertices of the dark lattice, $Z$ stabilisers are on faces and data qubits are on edges. Ancilla qubits used for stabiliser measurement are shown in dark and light for $X$ and $Z$ stabilisers respectively. Data qubits are shown in red. Weight three implementations of $\overline{X}$ and $\overline{Z}$ are also shown. (b) A 3D surface code. Qubits are placed on edges as in the 2D code, but to improve readability these qubits are only shown explicitly on the bottom and back-right boundaries.} 
    \label{fig:kitaev}
\end{figure}

We start by considering the 2D surface code~\cite{kitaev2003,bravyi1998} as in Fig.~\ref{fig:kitaev}. Here we are using the ``Kitaev picture'' where (with respect to the dark lattice) data qubits (red) are associated with edges. To each vertex of this lattice we associate a stabiliser generator $X(v) = \prod_{\{e | v \in e\}}X_e$ which acts with Pauli $X$ on all edges $e$ that meet at vertex $v$. To each face of this lattice we associate a stabiliser generator $Z(f) = \prod_{e \in f} Z_e$ which acts on all edges belonging to face $f$. Ancilla qubits used in measurement of these operators are also shown in Fig.~\ref{fig:kitaev} (dark for $X$ and light for $Z$). Also shown in Fig.~\ref{fig:kitaev} (in light grey) is the dual lattice, obtained by replacing the faces of the original lattice with vertices and the vertices with faces. In this lattice the Z stabilisers are on vertices and the X stabilisers are on faces. 

There are two types of boundary in this code which are commonly referred to as rough (left and right) and smooth (top and bottom). Logical $X$ operators are strings of single-qubit $X$ operators running between smooth boundaries while logical $Z$ operators are strings of single-qubit $Z$s running between rough boundaries. As such, we will henceforth refer to smooth and rough boundaries as $X$ and $Z$ boundaries respectively. 

Fig.~\ref{fig:kitaev} shows how to obtain a 3D surface code~\cite{dennis2002} from a 2D one, simply by extending the lattice into the third dimension. The dark square lattice has become a cubic one but we have retained the same assignment of parts of the code to geometric features of the lattice, i.e.\ data qubits are on edges, $X$ stabilisers are on vertices and $Z$ stabilisers are on faces. Four of the boundaries (top, bottom, front-left and back-right) of the code are $X$ boundaries and resemble 2D surface codes. $\overline{X}$ in this code is a sheet of single-qubit $X$ operators running between all four of these boundaries. The other two boundaries are $Z$ boundaries and $\overline{Z}$ is a string of single-qubit $Z$s running between these boundaries. 

To see why the dimension of $\overline{X}$ has changed but the dimension of $\overline{Z}$ has not we can look at the structure of the stabiliser generators. In the 2D surface code each qubit is part of at most two $X$ and two $Z$ stabiliser generators so single-qubit errors of either type only violate a pair of generators. Longer strings of $X$ or $Z$ errors will commute with all generators along their length and only anticommute with a single generator at each end of the string. A logical operator corresponds to a string with both of its endpoints connected to the relevant boundaries where they cannot be detected. In 3D, as in 2D, every qubit is part of at most two $X$ generators and so logical $Z$ operators are once again strings. However, in the bulk of the cubic lattice four faces meet at every edge so qubits in the bulk are part of four $Z$ generators. This means that a string of $X$ errors will be detected not just by the generators at its endpoints but also by generators adjacent to the string along its entire length. In other words, the $Z$ stabiliser syndromes are loops around regions containing $X$ errors. This means that instead of $\overline{X}$ being a string with its endpoints connected to boundaries it must be a membrane with its entire perimeter connected to boundaries.

These loop-like $Z$ syndromes have another effect: they allow us to detect measurement errors. We know that valid syndromes should form closed loops, so syndromes not satisfying this property must have been produced (at least in part) by measurement errors. This allows us to repair these syndromes (by joining the endpoints of strings to form loops) and removes the need for repeated measurements of these stabilisers in a process termed \textit{single-shot error correction}~\cite{bombin_single-shot_2015, campbell_theory_2019}. This stands in contrast to the 2D case where measurement errors can only be detected by repetition of these measurements.

\begin{figure}
    \begin{subfigure}{.3\textwidth}
        \includegraphics[width=.9\textwidth]{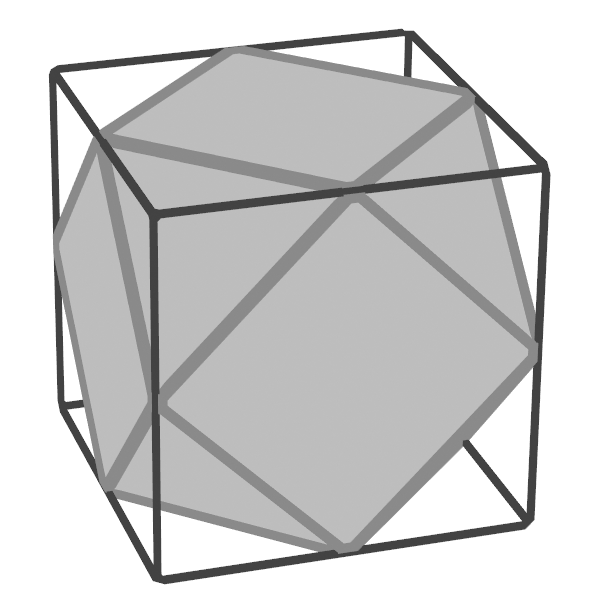}
        \subcaption{}
    \end{subfigure}
    ~~~
    \begin{subfigure}{.3\textwidth}
        \includegraphics[width=.9\textwidth]{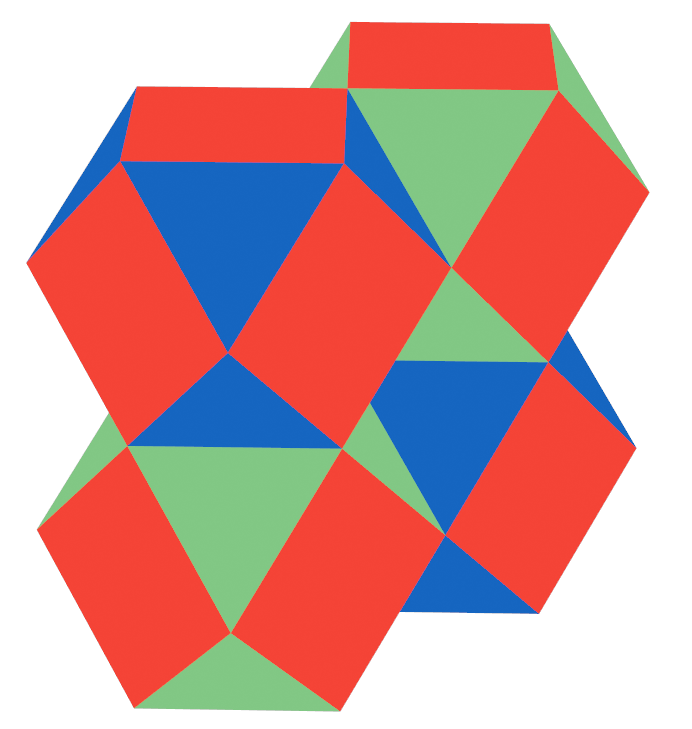}
        \subcaption{}
    \end{subfigure}
    ~~~ 
    \begin{subfigure}{.3\textwidth}
        \includegraphics[width=\textwidth]{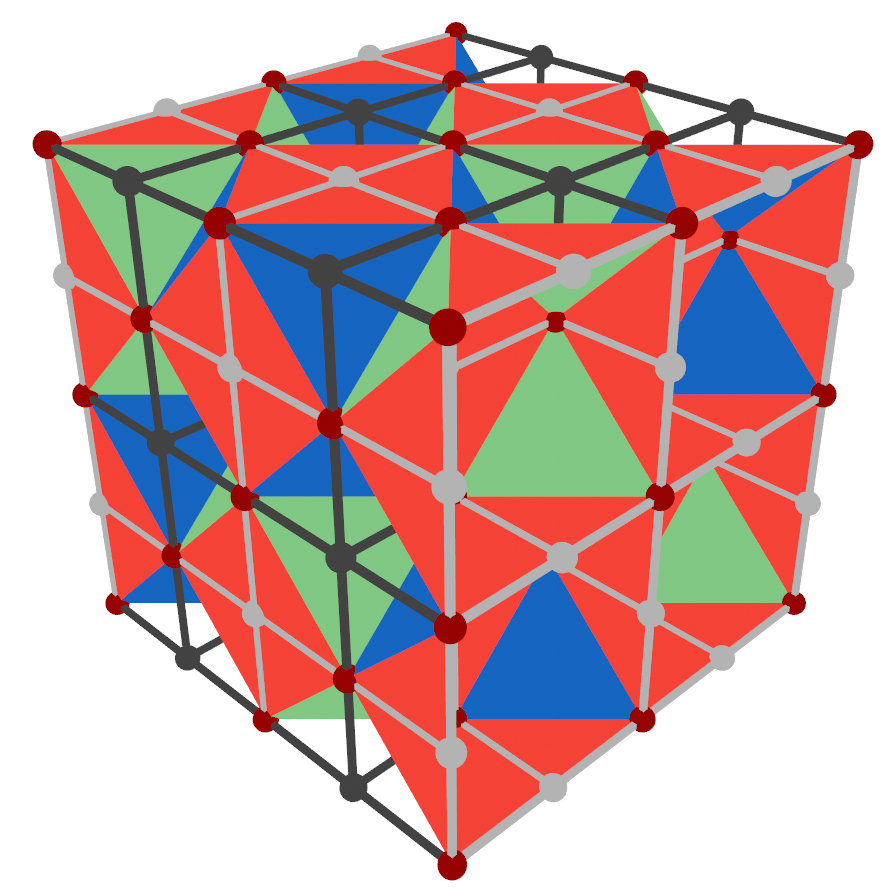}
        \subcaption{}
    \end{subfigure}
    \caption{(a) Original cubic lattice (dark edges) and cuboctahedral cell of rectified lattice (light faces and edges). There will also be one octahedral cell around each vertex of the original lattice. (b) Four cuboctahedral cells of the rectified lattice. Half of an octahedral cell formed from the negative space of the cuboctahedra can be seen at the centre. Square faces are coloured red and triangular faces are coloured green or blue such that each cell only has two different colours of face. (c) Correspondence between distance-3 surface codes in the Kitaev (as shown in Fig.~\ref{fig:kitaev}) and rectified pictures. A correspondence can be seen between $Z$ stabilisers (faces of dark cubic lattice) in the Kitaev picture and red faces of the rectified lattice. There is also a correspondence between $X$ stabilisers (vertices) of the cubic lattice and octahedra in the rectified lattice (which contain no red faces).}
    \label{fig:rectified}
\end{figure}

We will use the rectified lattice picture of 3D surface codes to construct our slices since it allows us to describe all three codes using the same lattice~\cite{vasmer_three-dimensional_2019}. This lattice is obtained from the standard cubic lattice by adding a new vertex at the middle of each edge, connecting these vertices if they belong to the same face and then deleting the original lattice. This results in one cuboctahedral cell per cubic cell of the original lattice (shown in Fig.~\ref{fig:rectified} (a)) and the negative space between these cells produces additional octahedral cells (half of such a cell can be seen at the centre of Fig.~\ref{fig:rectified} (b)). Fig.~\ref{fig:rectified} (b) also shows a colouring of the rectified lattice where each cell has two colours of face (red-blue and red-green cuboctahedra are visible and octahedral cells will be blue-green). It is then possible to simultaneously define three 3D surface codes on this lattice: three qubits are associated with each vertex and one colour is associated with each code. $c$-faces (for $c \in \{r,g,b\}$) represent $Z$ stabilisers in the $c$ code and cells containing no $c$-faces represent $X$ stabilisers. An example is shown in Fig.~\ref{fig:rectified} (c).

\section{Dimension Jumping in Surface Codes}
\label{section:dimension_jumping}
The ability to swap between a thin slice of 3D surface code and a 2D surface code (a process termed ``dimension jumping~\cite{bombin_dimensional_2016}'') is at the heart of Brown's procedure. Similarly to JIT decoding, this process was originally studied by Bomb\'{i}n for use in the 3D tetrahedral colour code and was adapted for use in the surface code by Brown~\cite{brown_fault-tolerant_2020}, although a process equivalent to the 3D $\rightarrow$ 2D collapse was previously studied by Raussendorf, Bravyi and Harrington in a measurement-based setting~\cite{raussendorf_long-range_2005}. Despite this, there is (to our knowledge) no comprehensive explanation of dimension jumping in surface codes in the literature, so we use this section to provide one. 

\subsection{2D to 3D expansion}
\label{subsection:d_j1}

The expansion process as originally presented (for the colour code) consists of the following steps:

\begin{itemize}
    \item Start with a 2D code, which is a boundary of a 3D code.
    \item Measure stabiliser generators of the 3D code.
    \item Apply error correction to the 3D code.
\end{itemize}

However, unlike the 3D colour code the 3D surface code minus a boundary still encodes a qubit and so we must take extra steps to ensure that the final logical state of the 3D code is the same as the initial logical state of the 2D code. Additionally, in the specific case where at each timestep we are combining dimension jumping with the application of $CCZ$ gates to part of the code we want our dimension jump to involve only measurements of $Z$ stabilisers. This is because the $X$ stabilisers do not all commute with this application of $CCZ$ gates and so the 2D codes at intermediate steps of the procedure will not be in eigenstates of these $X$ stabilisers. As such, measuring these stabilisers will change the state of the code and the overall effect of the procedure will not be a logical $CCZ$. 

\begin{figure}
    \begin{subfigure}{.33\textwidth}
        \includegraphics[width=\textwidth]{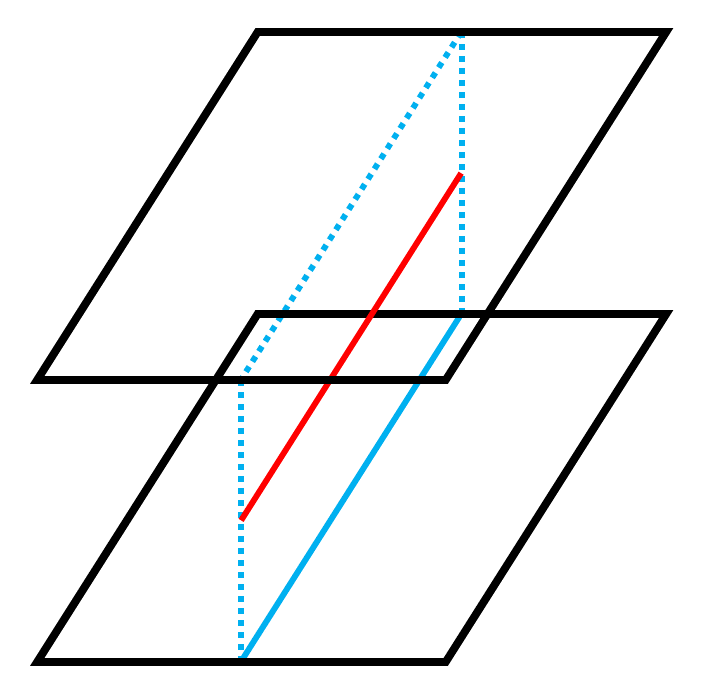}
        \subcaption{}
    \end{subfigure}
    \begin{subfigure}{.33\textwidth}
        \includegraphics[width=\textwidth]{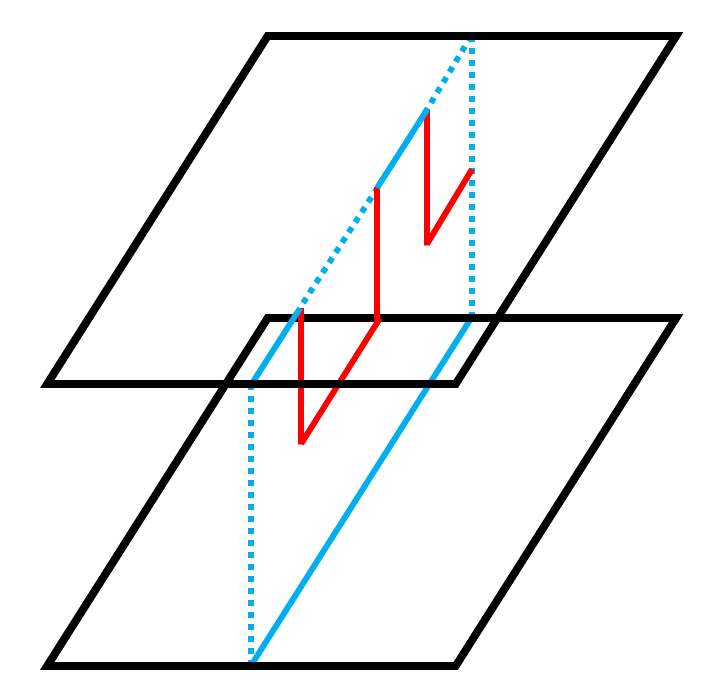}
        \subcaption{}
    \end{subfigure}
    \begin{subfigure}{.33\textwidth}
        \includegraphics[width=\textwidth]{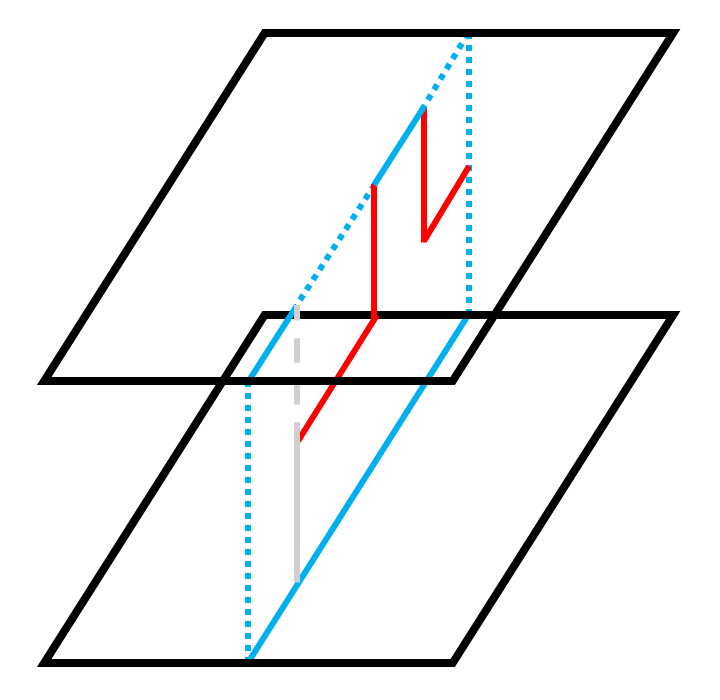}
        \subcaption{}
    \end{subfigure}
    \caption{(a) Two 2D surface codes (black) which we imagine are entangled by measurement of intermediate stabilisers (not shown) to form a slice of 3D surface code. We imagine that we have applied a logical $\overline{X}$ to the lower code (solid blue line) but not to the upper code. This results in the red syndrome on the intermediate stabilisers\protect\footnotemark. To successfully transfer the state of the lower 2D code into the 3D code we must apply a matching 2D logical $X$ to the top code which completes the logical $X$ of the 3D code (dashed blue). We can equivalently think of this as applying a correction which pushes the syndrome onto the top layer of the code. (b) In this example the the top code contains some strings of $X$ errors. The red syndrome now consists of loops joined to the top and side boundaries and ``filling in'' these loops to push them to the top boundary will once again correctly transfer the state of the lower 2D code to the 3D one. (c) Two measurement errors on a top code (dashed grey) and bottom code (solid grey) stabiliser cause us to lose track of what is inside and outside of a loop, making it difficult to reliably transfer the state of the 2D code into the bulk (the stabilisers are in the plane of the 2D code but the corresponding syndromes are perpendicular).} 
    \label{fig:transfer}
\end{figure}

A suitable dimension jump is implemented by 

\begin{itemize}
    \item Start with a 3D surface code $\mathcal{C}$ and its boundary $\partial \mathcal{C}$, which is a 2D surface code in a state $\ket{\overline{\psi}}$. These codes must be chosen such that the $Z$ stabilisers of the 2D code commute with the $X$ stabilisers of both the 2D and 3D codes. 
    \item Prepare all qubits belonging to $\mathcal{C} \setminus \partial \mathcal{C}$ in $\ket{+}$.
    \item Measure the $Z$ stabiliser generators of the 3D code that are not $Z$ stabilisers of the 2D code.
    \item Apply a correction which returns the 3D code to its codespace. This correction should not have support on any of the qubits of the original 2D code.
\end{itemize}

The preparation of new qubits in $\ket{+}$ combined with the commutivity requirement on the $Z$ stabilisers in the 2D code ensures that measurement of the 3D $Z$ stabilisers will project the encoded state either to a state in the codespace of the 3D code or to a state that can be returned to the codespace using a correction inferred from the measurement outcomes of these stabilisers. The distribution of $X$ errors that this correction addresses are what is referred to by Bomb\'{i}n as a Pauli frame~\cite{bombin_2d_2018} and by Brown as a gauge of the 3D code~\cite{brown_fault-tolerant_2020}. 

To understand why is it important not to measure the $Z$ stabilisers of the 2D code during the dimension jump and why the correction should have no support on this code it is easiest to consider an idealised example as in Fig.~\ref{fig:transfer}. In (a) and (b) we see how applying corrections to the top layer stretches the string-like logical $\overline{X}$ of the lower 2D code into the sheetlike logical $\overline{X}$ of the 3D code and in (c) we see the issues that can arise if we allow for measurement of stabilisers of the initial 2D code during the jump (and thus allow for the possibility of measurement errors on these stabilisers). Note that in Fig.~\ref{fig:transfer} we begin with two surface codes and then entangle them via intermediate stabiliser measurements instead of initialising and entangling the top code in a single step by measuring top and intermediate stabilisers simultaneously. These two cases are equivalent since stabiliser measurements commute by definition. We do not need to do anything to ensure the correct transfer of $\overline{Z}$ from the 2D code into the 3D one because $\overline{Z}$ of either 2D code is a valid implementation of $\overline{Z}$ in the 3D code. By preparing the top code in $\ket{\overline{+}}$ (by preparing the physical qubits in $\ket{+}$ and measuring $Z$ stabilisers), we can ensure no logical $Z$ is applied in this code, so any implementation of $\overline{Z}$ applied to lower 2D code will be equivalently applied to the 3D code. 

\footnotetext{This syndrome occurs because the intermediate $Z$ stabilisers do not commute with the 2D $X$ logical. However, they also do not commute with the 2D $X$ stabilisers and so in reality we would expect to observe this boundary-to-boundary syndrome string (from the logical) plus a random pattern of syndrome loops (from the stabilisers). This does not change any of the analysis however, so it is sufficient to consider only the zero-loop case.}

\subsection{3D to 2D collapse}
\label{subsection:d_j2}

\begin{figure}
    \begin{minipage}{.35\textwidth}
        \begin{subfigure}{.9\textwidth}
            \includegraphics[width=.8\textwidth]{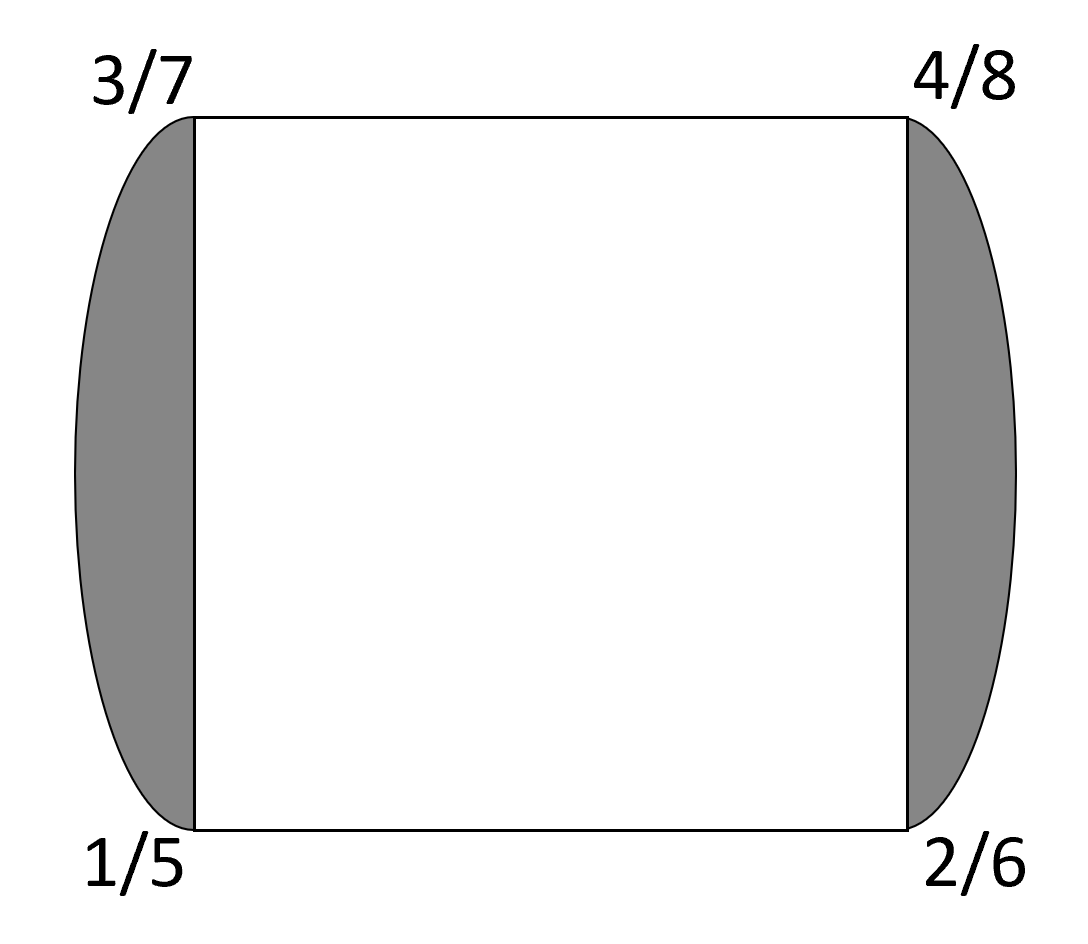}
            \subcaption{}
        \end{subfigure}
        \\
        \begin{subfigure}{.9\textwidth}
            \includegraphics[width=.8\textwidth]{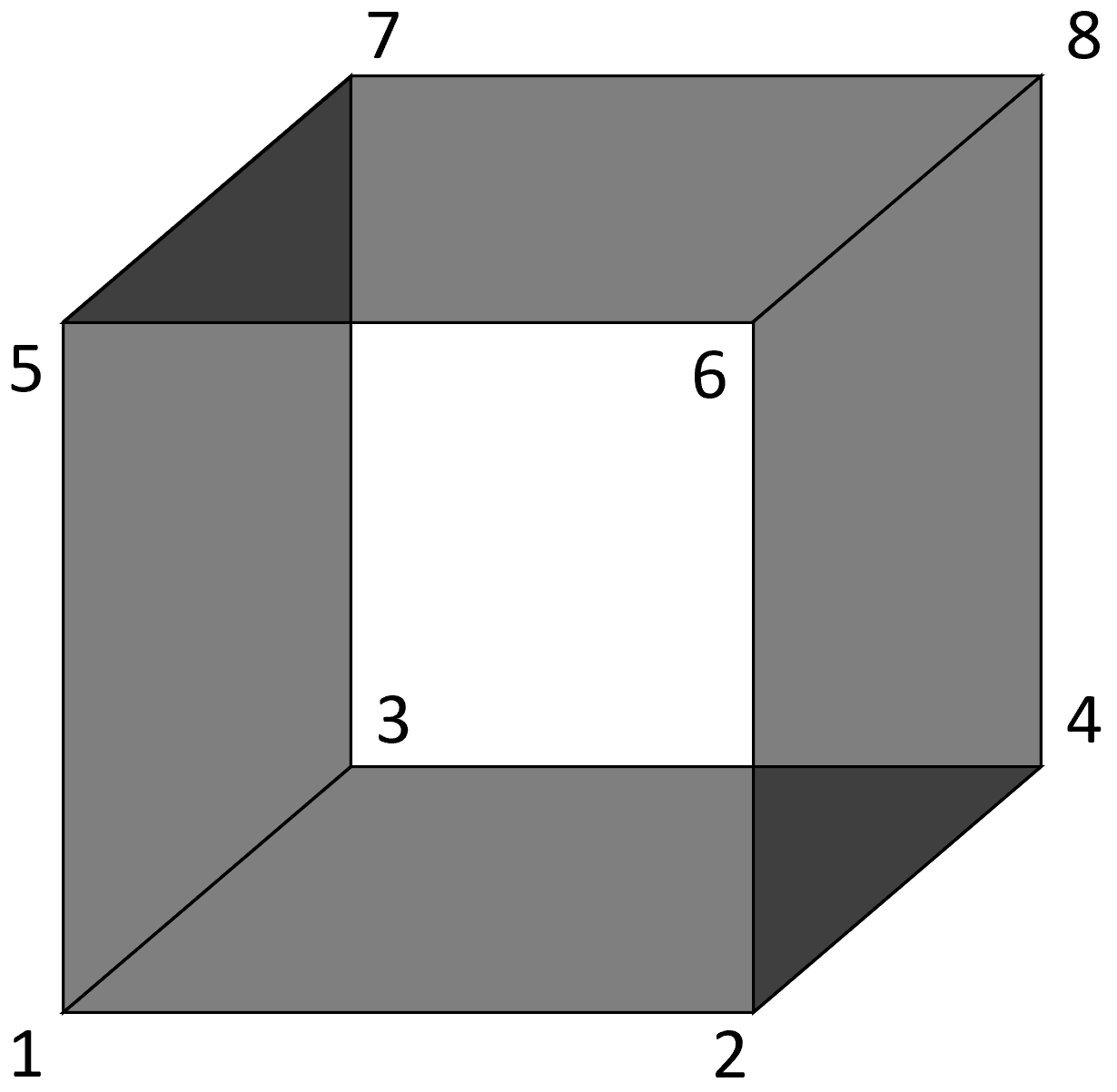}
            \subcaption{}
        \end{subfigure}
    \end{minipage}
    \begin{minipage}{.65\textwidth}
        \caption{2x2 and 2x2x2 minimal surface codes. Qubits are on vertices. (a) has two $X$ stabiliser generators (dark boundaries) and one $Z$ stabiliser (light face) so encodes one qubit. In (b) there is a $Z$ stabiliser on each face of the cube and an $X$ stabiliser on each of the dark boundaries. Only three of these $X$ and four of these $Z$ stabilisers are independent so this code also encodes a single qubit. A version of (a) exists on the top and bottom faces of (b). An implementation of $\overline{X}$ is supported on either of the light faces of the 3D code and on the top and bottom edges of the 2D code while $\overline{Z}$ is supported on either of the other two edges in the 2D code and on these same edges in the 3D code.}
        \label{fig:2x2x2}
    \end{minipage}
\end{figure}

The collapse part of dimensional jumping can similarly be adapted to surface codes. To understand how the process works in this case it is easiest to consider a simple example, namely the one given in Fig.~\ref{fig:2x2x2}. Here we have minimal examples of (a) a 2x2 and (b) a 2x2x2 surface code. We can switch between these two codes using dimension jumping since a version of (a) exists on the top and bottom faces of (b). 
A generic state for this code can be written 

\begin{equation}
    \ket{\overline{\psi}} = a\ket{\overline{+}} + b\ket{\overline{-}}
\end{equation}

\noindent and we can write $\ket{\overline{+}}$ and $\ket{\overline{-}}$ as 

\begin{equation}
    \ket{\overline{+}} = \frac{1}{\sqrt{n}}\sum_i^n Z_i\ket{{+}{+}...+}_{1-8},
    ~~~
    \ket{\overline{-}} = \frac{1}{\sqrt{n}}\sum_i^n \overline{Z}Z_i\ket{{+}{+}...+}_{1-8}
\end{equation}

\noindent where $Z_i$ are stabilisers of the code and $\overline{Z}$ is an implementation of logical $Z$. Now consider the case where we begin in a logical state of the 3D code and measure qubits 1-4 in the $X$ basis. We can choose $\overline{Z}$ to be supported on qubits 5 and 7 or 6 and 8 so only the stabilisers $Z_i$ are supported on the bottom four qubits. The measurement projects to a state 

\begin{equation}
    \label{eq:proj_state}
    \ket{\phi} = \frac{\ket{{\pm}{\pm}{\pm}{\pm}}_{1-4}}{\sqrt{m}}\sum_j Z_j^m\ket{{+}{+}{+}{+}}_{5-8}
\end{equation}

\noindent where $Z_j$ are the restrictions of a subset of the operators $Z_i$ to the top four qubits, possibly composed with $\overline{Z}$. This subset is all those $Z_i$ which match the measurement outcomes from the bottom four qubits, i.e. their restriction to these qubits is supported only on qubits where we measured $-1$. If all four single-qubit measurement outcomes were the same then the only possible $Z_j$ are $Z_j = I$ and $Z_j = Z_5Z_6Z_7Z_8$ and the sum in (\ref{eq:proj_state}) is a state in the codespace of the 2D code. If the single-qubit measurement outcomes are not all the same then the 2D code contains errors. For instance, if measurement of qubits 1-4 projects these qubits to the state $\ket{{-}{-}{+}{+}}$ then the possible $Z_j$ are $Z_5Z_6$ and $Z_7Z_8$ which are both 2-qubit errors. To return to the codespace we can apply a correction $Z_5Z_6$, and more generally we apply $Z$ to one of the top qubits whenever we measure $-1$ from the corresponding bottom qubit. For codes which are larger or have more complicated geometries the process is not quite so straightforward but a correction can still be found by considering the stabiliser structure. 

In the case where there are $Z$ errors in the bottom code it is no longer possible to infer the necessary correction just from the measurement outcomes of the bottom face, but we can combine these measurements with measurements of the $X$ stabilisers of the 2D code to identify these errors (by reconstructing the syndrome of the 3D code). In our example, an odd parity of bottom face measurements tells us $X_1X_2X_3X_4$ would be violated, the parity of qubits 1 and 3 (2 and 4) together with the measurement outcome of $X_5X_7$ ($X_6X_8$) allows us to infer the measurement outcomes of the two side-face $X$ stabilisers and the product of $X_5X_7$ and $X_6X_8$ gives us the outcome of the top-face stabiliser. Because this code is only distance-2 we cannot correct for an error in this case, but in higher-distance codes we can.

An additional complication is introduced in the case where we apply $CCZ$ gates to a subsection of the 3D code during each timestep. In this case we cannot measure the $X$ stabilisers of the 2D code post-collapse (the state we project to in the collapse will not be an eigenstate of these stabilisers), and so we cannot fault-tolerantly infer corrections for the top face. Fortunately, this is not an issue because $Z$ errors commute with $CCZ$ and so applying a single $Z$ correction to the final code is equivalent to applying error correction throughout. This correction is obtained in two steps: firstly we must combine the $X$ stabiliser measurements from the final 2D code with single-qubit measurement outcomes from previous slices to obtain an $X$ stabiliser syndrome for a 3D surface code and decode this syndrome to identify the locations of $Z$ errors. Secondly, we combine these $Z$ error locations with the single-qubit measurement outcomes to obtain a corrected set of single-qubit measurement outcomes, and from these we calculate a correction for the final 2D code. This operator can be interpreted as the adaptive $Z$ correction of a teleportation circuit, and this perspective on the procedure is discussed further in~\cite{webster_universal_2022}.

\section{Constructing Slices}
\label{section:codes}
In this section we present our proposed slices through the three 3D surface codes. We begin with an examination of the criteria which these slices must satisfy, then discuss each of our slices individually and finally demonstrate that they have the correct overlap at each step of the procedure. 

\subsection{Criteria for Valid Slices}

We now examine some necessary criteria which slices through 3D codes must meet in order to be used in Brown's procedure. This list is by no means exhaustive, but serves to highlight some of the major issues which must be avoided when constructing slices. 

Firstly, we have the requirement that any representative of a logical operator in the slice must have weight at least $d$ in order to preserve the distance of the code. The ability to construct slices satisfying this relies on a specific property of the 3D surface code (the string-like logical operator of code is required to run between a particular pair of boundaries) which is not present in general 3D codes. This restriction on the string-like logical means that as long as we ensure that its associated boundaries are on the sides of the slice and not on the top and bottom we are guaranteed logical operators with weight at least $d$. It is this restriction that forces at least one of the three codes to have a different time direction to the other two, since the string-like logical operators of three 3D surface codes which admit a transversal $CCZ$ are all perpendicular~\cite{vasmer_three-dimensional_2019} so there is no way to define a consistent time direction for all three (in three dimensions) such that slices in all three codes have this property. 

Secondly, we require that all $Z$ stabilisers in the layers commute with all $X$ stabilisers in the slices. This requirement is due to a combination of the dimension jumping process and the $CCZ$ gate as discussed in Section~\ref{section:dimension_jumping}. In addition, we require that the support of a 3D $X$ stabiliser on a layer corresponds exactly to a 2D $X$ stabiliser within that layer. This is what ensures that measurements of $X$ stabilisers in the final layer alongside the single-qubit measurement history throughout the procedure can reproduce the full $X$ stabiliser syndrome of a 3D surface code. This rule prevents us from having things such as a 3D $X$ stabiliser which is supported on only a single qubit of a layer.

Finally, all three slices must agree on a common set of overlapping qubits at every timestep. Stated differently, this means that qubits between which we intend to apply $CCZ$ must all be live simultaneously. This is essential for the procedure or the three codes would disagree on which qubits $CCZ$ should be applied to at each timestep.

The structures originally proposed by Brown were ``staircases'' through the cubic lattice in the Kitaev picture. In this section we present an alternate set of lattice slices which we find easier to understand and simulate, although we believe equivalent simulations could be performed using the staircase slices. 

\subsection{Proposed Layers and Slices}

\begin{figure}
    \begin{subfigure}{.5\textwidth}
        \includegraphics[width=.9\textwidth]{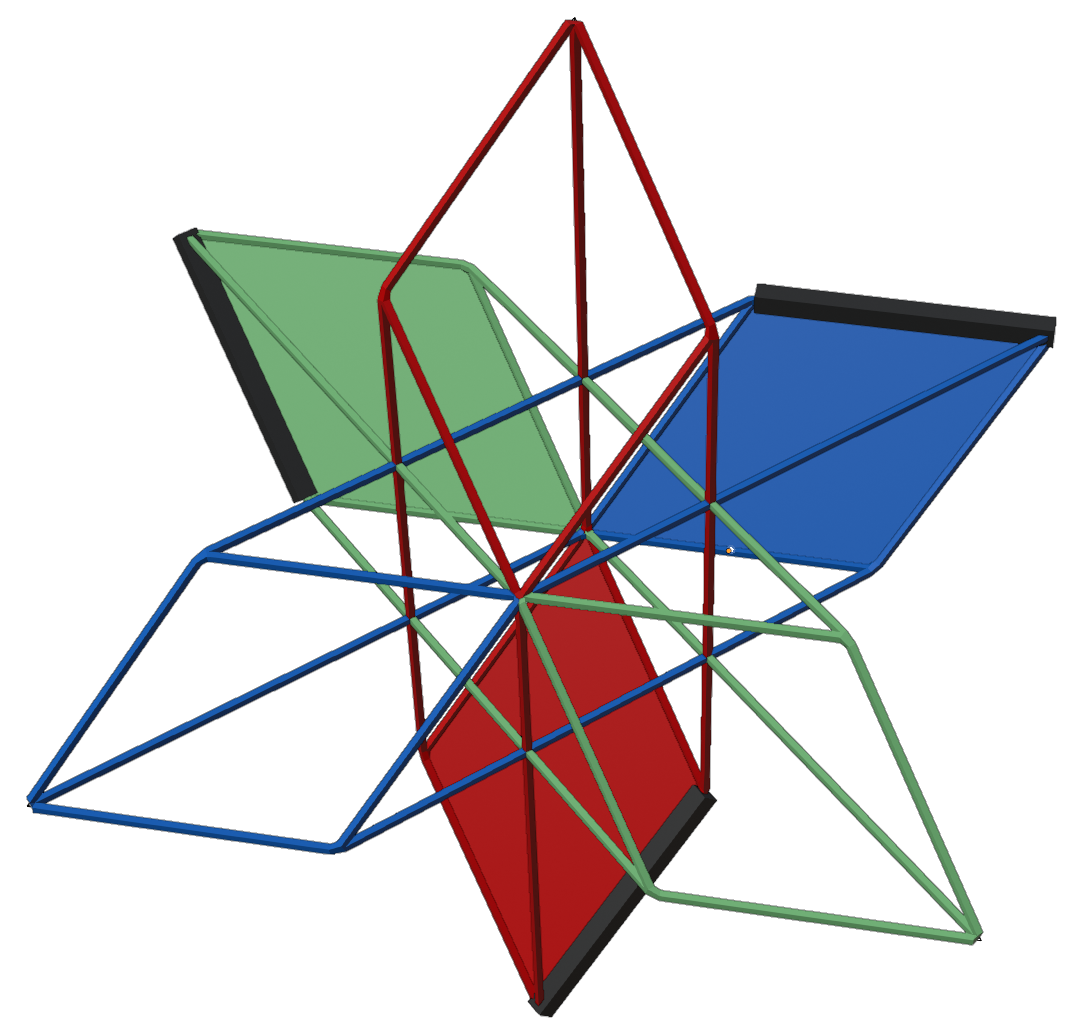}
        \subcaption{}
        \label{subfig:layersA}
    \end{subfigure}
    ~~
    \begin{subfigure}{.5\textwidth}
        \includegraphics[width=.9\textwidth]{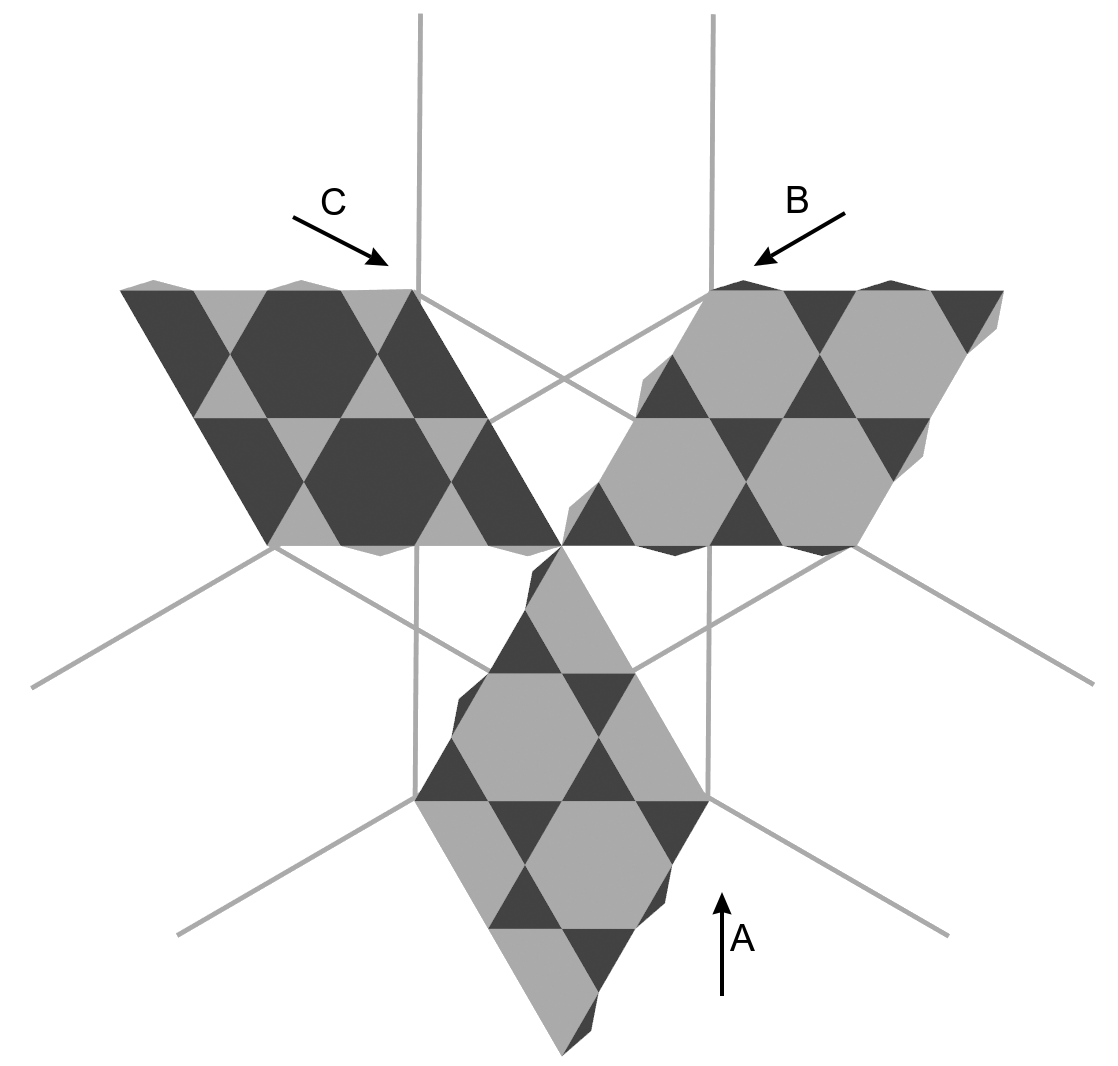}
        \subcaption{}
        \label{subfig:layersB}
    \end{subfigure}
    \caption{(a) 3D spacetime for all three surface codes. Each of the three 2D codes sweeps out a 3D surface code over time and the cubic region where all three of these codes intersect supports a transversal $\overline{CCZ}$. An animated version of this figure showing the motion of the 2D codes can be found at~\cite{gitrepo}. (b) Microscopic details of three distance-3 2D surface codes (light faces are $Z$ stabilisers and dark faces are $X$ stabilisers, qubits are on vertices) and and their directions of motion during the procedure. Logical $Z$ operators for the 2D layers are shown in black for each code in (a) and by comparing with (b) we can see that they run between $Z$ boundaries.}
    \label{fig:layers}
\end{figure}

The layers we propose are shown at a macroscopic and microscopic level (for distance 3) in Fig.~\ref{fig:layers} (a) and (b) respectively. In Brown's original proposal the time direction was the same for two of the codes and different for the third, but here we use a different time direction for each code. This is still implementable in 2D (also shown in Fig.~\ref{fig:layers} (b)). In the following subsections we show a distance-3 example of the corresponding slice through each of the three 3D codes and then discuss their overlap \footnote{Whilst we have endeavored to make the following figures as clear as possible, these codes are most easily understood using 3D models rather than 2D images. A .blend file containing various helpful models can be found at~\cite{gitrepo}}.

\begin{figure}
    \begin{minipage}{.4\textwidth}
        \begin{subfigure}{\textwidth}
            \includegraphics[width=\textwidth]{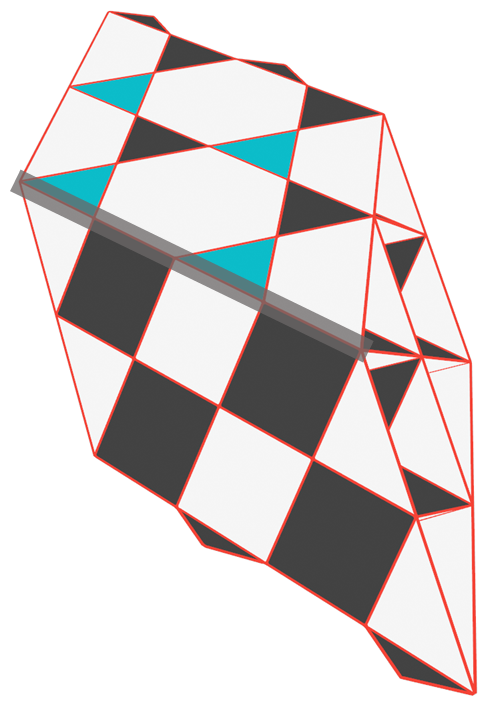}
            \subcaption{}
        \end{subfigure}
    \end{minipage}
    \begin{minipage}{.6\textwidth}
        \begin{subfigure}{\textwidth}
            \includegraphics[width=\textwidth]{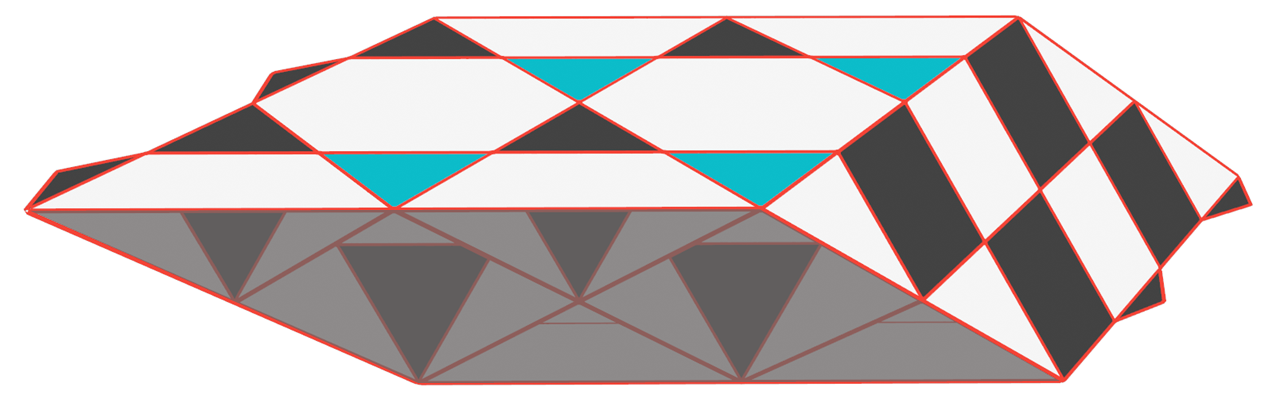}
            \subcaption{}
        \end{subfigure}
        \caption{A three layer thick slice through code A. $X$ stabilisers are dark cells and $Z$ stabilisers are light faces. Cyan faces are 2D $X$ stabilisers on $X$ boundaries. In (a) the front-right and back-left boundaries are $Z$ boundaries and the rest are $X$ boundaries. The hexagonal faces are not stabiliser generators but are each the product of three weight-4 stabilisers, while the half-hexagons are the product of two weight-3 stabilisers. A representative of logical Z/X is shown in grey in (a)/(b).}
        \label{fig:codeA}
    \end{minipage}
\end{figure}

\begin{figure}
    \begin{minipage}{.4\textwidth}
        \begin{subfigure}{\textwidth}
            \includegraphics[width=\textwidth]{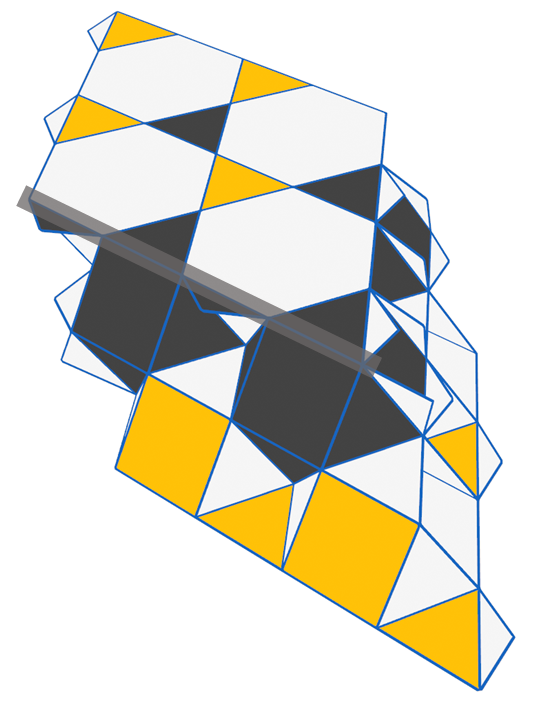}
            \subcaption{}
        \end{subfigure}
    \end{minipage}
    \begin{minipage}{.6\textwidth}
        \begin{subfigure}{\textwidth}
            \includegraphics[width=\textwidth]{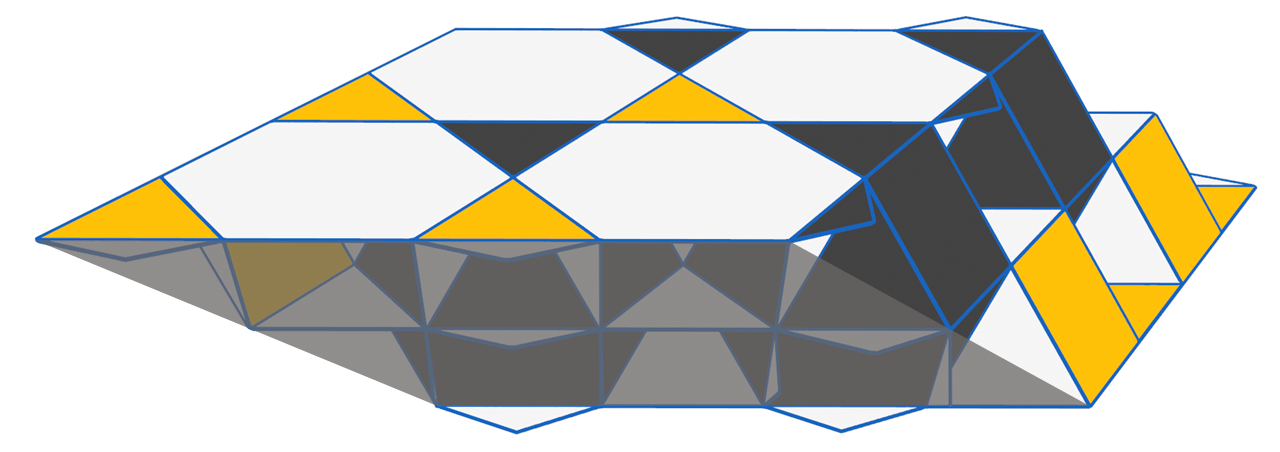}
            \subcaption{}
        \end{subfigure}
        \caption{A three layer thick slice through code B. $X$ stabilisers are dark cells and $Z$ stabilisers are light faces. In (a) the front-right and back-left boundaries are $Z$ boundaries and the rest are $X$ boundaries. Yellow faces are 2D $X$ stabilisers. Unlike in code A they are present on all 4 $X$ boundaries and there are also weight-2 $Z$ stabilisers on the $Z$ boundaries. The hexagonal faces are each the product of four weight-3 stabiliser generators. A representative of logical Z/X is shown in grey in (a)/(b).}
        \label{fig:codeB}
    \end{minipage}
\end{figure}

\begin{figure}
    \begin{minipage}{.4\textwidth}
        \begin{subfigure}{\textwidth}
            \includegraphics[width=\textwidth]{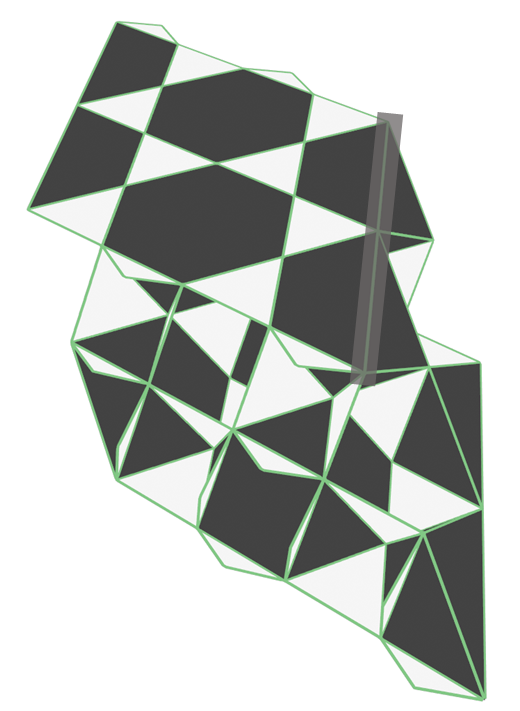}
            \subcaption{}
        \end{subfigure}
    \end{minipage}
    \begin{minipage}{.6\textwidth}
        \begin{subfigure}{\textwidth}
            \includegraphics[width=\textwidth]{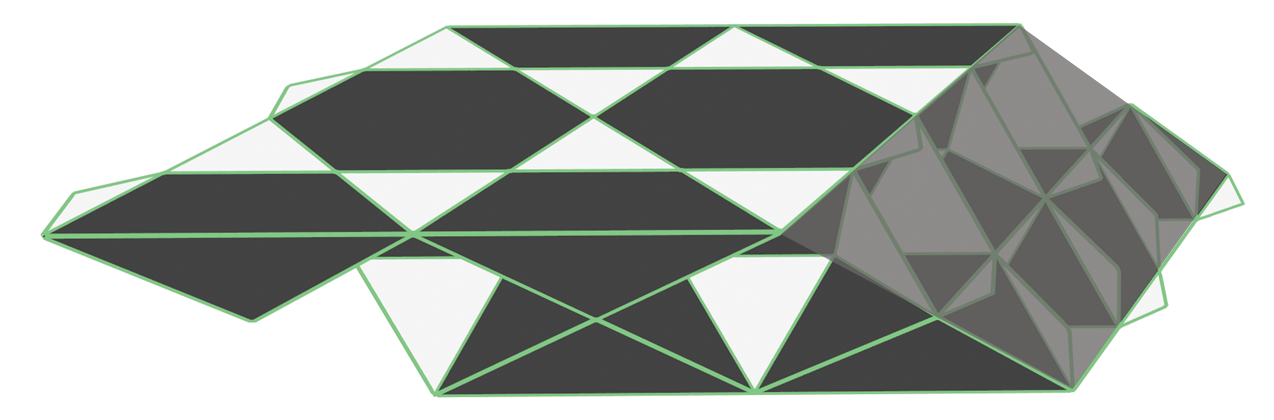}
            \subcaption{}
        \end{subfigure}
        \caption{A three layer thick slice through code C. $X$ stabilisers are dark cells and $Z$ stabilisers are light faces. In (a) the front-left and back-right boundaries are $Z$ boundaries and the rest are $X$ boundaries. As with code B there are weight-2 $Z$ stabilisers on the $Z$ boundaries but there are no 2D $X$ stabilisers on $X$ boundaries. Unlike both previous codes, the $Z$ stabilisers on the top and bottom boundaries are stabiliser generators of the 3D code but those of the middle layer are not. A representative of logical Z/X is shown in grey in (a)/(b).}
        \label{fig:codeC}
    \end{minipage}
\end{figure}

\subsubsection{Code A}

This is the simplest of the three codes and corresponds to the red code on the rectified lattice. As illustrated in Fig.~\ref{fig:rectified} (c) this is the standard 3D surface code defined on a simple cubic lattice. Examples of the boundaries are shown in Fig.~\ref{fig:codeA} for a distance-3 slice. This slice is three layers thick, but it is actually possible to define a two-layer-thick slice in this code since the middle layer in Fig.~\ref{fig:codeA} is identical to the top and bottom layers. This is not true for the other two codes, and so we must also use a three-layer-thick slice here to ensure the correct overlap of the three slices.

\subsubsection{Code B}

Code B is the blue code on the rectified lattice and a slice through this code is shown in Fig.~\ref{fig:codeB}. The $Z$ stabilisers of this code are weight-3 (in contrast to the weight-4 $Z$ stabilisers of code A) while its $X$ stabilisers are weight-12 (in code A they were weight 6). Because the lattice constant for the blue code is twice that of the red code we must use a slice that is three layers thick.

It is interesting to note that the top 2D code for this slice differs from that of code A only by a half-hexagon translation on the kagome lattice and so one might imagine that it would be possible to define a slice in code A with top and bottom 2D codes matching those seen in code B. However, the weight-2 $Z$ boundary stabilisers present in these 2D codes do not commute with the 3D $X$ stabilisers of code A and so it is not possible to expand from the 2D code to the 3D one by only measuring $Z$ stabilisers. 

\subsubsection{Code C}

Code C is the green code on the rectified lattice and a slice through this code is shown in Fig.~\ref{fig:codeC}. The top and bottom layers in this slice resemble the middle layer from the slice for code B, with $Z$ stabilisers on triangles and $X$ stabilisers on hexagons. The same is true for the middle layer of this slice and the top and bottom layers of the slice through code B. Additionally, these 2D codes are translated by a half-hexagon on the kagome lattice relative to those of code B and so have boundaries which match those of the 2D codes from code A. The inversion of stabilisers in these 2D codes relative to codes A and B mean that the weights of the logical operators are also exchanged (2D codes A and B had weight 3 $\overline{X}$ and weight 5 $\overline{Z}$ whereas here it is the opposite). Due to this, we observe better performance during JIT decoding in this code than in codes A and B (Fig \ref{fig:threshold_plots}), since the JIT decoder only deals with $X$ errors and Z errors are dealt with separately.  

\subsection{Overlap of the Three Codes}

\begin{figure}
    \centering
    \includegraphics[width=.8\textwidth]{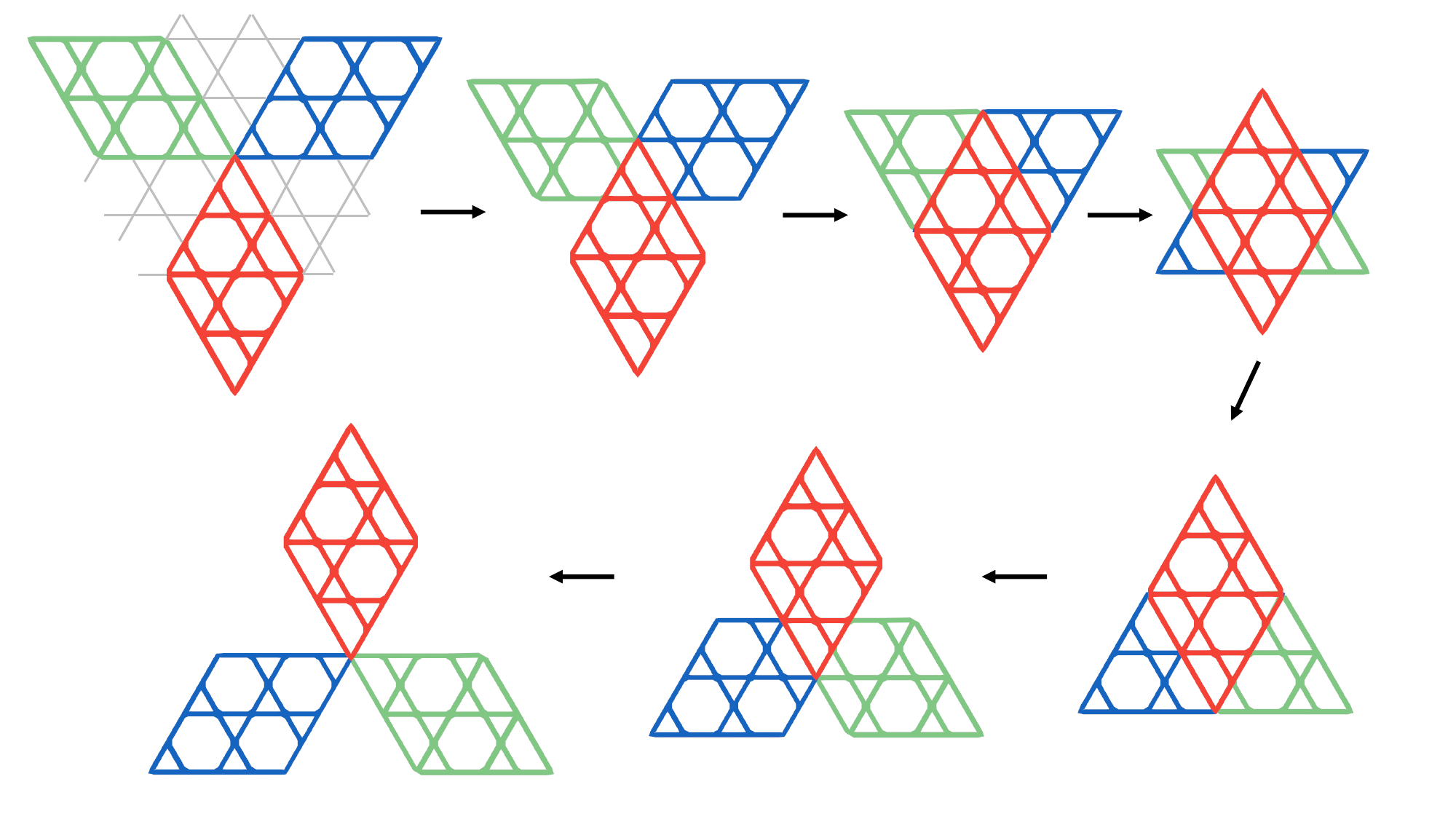}
    \caption{The three distance-3 2D surface codes from Fig.~\ref{fig:layers} (b) passing through each other. The vertices (qubits) of a given code always coincide with the vertices of the other two codes in the overlapping region. A single, global kagome lattice can be consistently overlaid on all three codes at all points of the procedure, and the three individual codes correspond to sections cuts from this lattice, parts of which are shown in grey in the first image.} 
    \label{fig:sequence}
\end{figure}

The full sequence of the distance-3 2D codes passing through each other can be seen in Fig.~\ref{fig:sequence}. Vertices of all three codes coincide in the overlapping region, and in fact, these three codes can all be thought of as sections cut from the same lattice. If our slices were two layers thick then our three 2D codes would follow this sequence exactly, but because they are three layers thick we actually move two steps in this sequence for every step of the procedure, i.e. if we start with the first code configuration then after one cycle of expand $\rightarrow$ correct $\rightarrow$ $CCZ$ $\rightarrow$ collapse we will have the third configuration. While we do not observe the second configuration in Fig.~\ref{fig:sequence} in a purely 2D setting it will correspond to the overlap of the three middle layers of the slices during the aforementioned set of operations. All qubits in each slice belong to one of the three layers and so all three layers agree on common sets of overlapping qubits.

\subsection{Practical Implementation}

This procedure would not be of much use if it were necessary to sweep through a physical 3D code. In Brown's proposal the time direction is changed at each timestep so that the physical architecture only needs to be one slice thick and we can move up and down between the top and bottom layers. Our slices are compatible with this, although the architecture must accommodate three layers instead of two. Additionally there must be room for two of the codes to move relative to the third (in the original proposal only one code needed to move) and this will increase the qubit cost of the procedure.

\section{Linear-Time CCZ}
\label{section:CCZ}
Three 3D surface codes defined on a $d \times d \times d$ cube in the rectified lattice with boundaries as in Fig.~\ref{fig:rectified} admit a transversal $\overline{CCZ}$ gate~\cite{vasmer_three-dimensional_2019}. Key to Brown's procedure is the idea that this gate is also implementable (in time linear in the code distance $d$) on three 2D surface codes whose overlap in spacetime is equivalent to such a 3D surface code. More precisely, as the three codes move through each other (via the dimension jumping process described in previous sections) $CCZ$ gates are applied between all overlapping triples of physical qubits that do not belong to the top layer. The exclusion of top-layer qubits is necessary because these will become the bottom-layer qubits of the next slice and we only want to apply $CCZ$ once to each qubit in the overlapping region. The intuition for why this process should work is that JIT decoding allows us to fault-tolerantly exchange a spatial dimension for a temporal one, so any process which gives a logical operation in $(3+0)$ dimensions should result in an equivalent logical operation in $(2+1)$ dimensions. However, some readers may not be satisfied with this argument and so it is both interesting and worthwhile to examine the workings of this gate in more detail.

To begin with, we neglect JIT decoding entirely and just consider the full 3D spacetime as shown in Fig.~\ref{subfig:layersA}. The first thing to note about the spacetime of the three codes is that it is not a $d \times d \times d$ cube; rather, it is three parallelepipeds whose intersection is a $d \times d \times d$ cube. It is not immediately obvious, and nor is it generally true, that applying $CCZ$ transversally in this overlapping region should implement a logical $\overline{CCZ}$ between the codes. Recall that $\overline{CCZ}$ maps $\overline{X}_i$ to $\overline{X}_i\overline{CZ}_{jk}$ (where $CZ = 1/2(II + IZ + ZI - ZZ)$) and thus the intersection of $\overline{X}$ in any code with the region where we apply $CCZ$s must be a region supporting valid implementations of $\overline{Z}$ in the other two codes. In a version of \ref{subfig:layersA} where $\overline{Z}$ in each of the three codes runs in the ``long'' direction, the transversal application of $CCZ$ between the overlapping qubits in the central cubic region will not implement $\overline{CCZ}$ because there is no implementation of $\overline{Z}$ in any of the three codes which is fully supported inside the cube. Fortunately this is not the case for the arrangement of codes described in the previous section, where instead $\overline{Z}$ runs in one of the ``short'' directions in each code (as shown by the dark lines in Fig.~\ref{subfig:layersA}) and so each code possesses valid implementations of this operator which are contained within the cube. In contrast, there is no implementation of $\overline{X}$ contained within the cube in any of the codes, but this is not a problem since nothing is mapped to $\overline{X}$ by $\overline{CCZ}$. We require only that the intersection of $\overline{X}_i$ with the cube supports valid implementations of $\overline{Z}_j$ and $\overline{Z}_k$ and this requirement is satisfied. This means that a constant-time version of the procedure where we start with our initial set of 2D codes, use dimension jumping to expand to the full 3D spacetime, apply $CCZ$s to the overlapping region then collapse to the final set of 2D codes will implement the desired logical operation. 

Now that we are satisfied that a single-timestep version of the procedure implements $\overline{CCZ}$ we can convince ourselves of the same thing for the linear-time version using induction. Consider the following two sets of operations, where ``expand'' should be understood to mean ``measure 3D Z stabilisers then apply JIT decoding and $X$ error correction'' and we assume that these correction operations are successful:

\begin{itemize}
    \item \textbf{A:} Begin with a layer at position $x$, expand to a slice with thickness $\Delta x$, apply $CCZ$s, collapse to a layer at position $x + \Delta x$, then repeat to end at a layer at position $x + 2\Delta x$. 
    \item \textbf{B:} Begin with a layer at position $x$, expand to a slice with thickness $2\Delta x$, apply $CCZ$s and collapse to a layer at position $x + 2\Delta x$.
\end{itemize}

Each qubit experiences the same set of operations in both cases: measurement of associated $Z$ stabilisers and an $X$ correction if one is required (for qubits not in the initial layer), application of $CCZ$ (for qubits in the overlapping region and not in the final layer) and measurement in the $X$ basis (for qubits not in the final layer). The only way for this sequence of operations to be inequivalent in \textbf{A} and \textbf{B} is if operations in the first slice in \textbf{A} influence our choice of $X$ correction for qubits in the second slice in a way that does not occur in \textbf{B}. Incorrect choices of $X$ error correction operator in the first slice of $\textbf{A}$ will have this effect but we are assuming that all correction attempts are successful\footnote{This will not be true in reality so this argument does not guarantee a threshold for the procedure. However, we are currently only asking if the gate has the correct logical action in principle and so these assumptions are justified.}. The only other operation applied to the qubits of the first slice of $\textbf{A}$ is $CCZ$, but this is not applied to qubits that are part of the second slice and its only effect will be to change the outcomes of our single-qubit measurements. As discussed in Section~\ref{section:dimension_jumping}, the correction inferred from these measurements is only applied at the very end of the procedure so we conclude that none of the operations on the first slice of $\textbf{A}$ can influence the operations on the second and the logical effects of \textbf{A} and \textbf{B} must be equivalent. This conclusion, combined with the knowledge that performing the entire procedure in a single timestep correctly implements $\overline{CCZ}$, allows us to infer that the linear-time version of the gate also implements $\overline{CCZ}$.

\section{The Delayed Matching Decoder}
\label{section:decoder}
\subsection{Description}

Now that we have constructed a set of slices we are ready to consider explicit error correction operations within these slices. We imagine that we are at a point of the procedure where we have prepared the new qubits and measured the new stabilisers but have not yet applied any corrections. We have not remeasured the bottom-layer stabilisers as we assume that our choice of correction in the previous timestep was successful and all of these stabilisers are in the +1 eigenstate. The $Z$ stabiliser measurements result in a random distribution of $X$ errors on the new qubits and in a syndrome for these errors consisting of a set of loops in the bulk or connected to the top or side $X$ boundaries. If there were errors on the bottom layer of qubits or if we had errors in some of our stabiliser measurements we will also have broken strings with endpoints which must be matched up to produce a valid syndrome. The delayed matching decoder is a modified version of the minimum-weight perfect matching (MWPM) decoder~\cite{dennis2002,Edmonds1965,fowler2014} and provides a simple but fault-tolerant method of performing this matching. This decoder was first proposed by Brown in~\cite{brown_fault-tolerant_2020} and its operation is is as follows:

\textbf{Setting:} We consider $Z$ stabilisers on faces and syndromes corresponding to (possibly broken) loops on edges dual to these faces\footnote{We recall that the dual of a 3D lattice is obtained by placing new vertices at the centres of cells of the original lattice, connecting these vertices if their cells share a face in the original lattice and then deleting the original lattice. This transformation maps cells to vertices, faces to edges and vice versa.}. The distance between endpoints of broken loops is measured in terms of path length on the dual edges connecting them\footnote{Some readers may prefer to think in terms of the qubits-on-faces picture where $Z$ stabilisers are on edges and endpoints are on vertices. For code A the relevant lattice is the simple cubic lattice, whereas for codes B and C it is the rhombic dodecahedral lattice. 3D models showing the relation between the previously defined slices in the qubits-on-vertices, qubits-on-edges and qubits-on-faces picture can be found at~\cite{gitrepo}.}. When we speak of joining pairs of endpoints to each other or to a boundary we mean that the stabiliser measurement outcomes are flipped (in software) along a path connecting this pair of objects. 

\textbf{Inputs:} 
\vspace{-5pt}
\begin{itemize}
    \setlength\itemsep{-0.5em}
    \item A vector $e$ of endpoint locations within the current slice 
    \item A map (i.e. a set of key-value pairs) $M$ from pairs of endpoint locations (or an endpoint location and a boundary) to a ``pseudodistance'' between the elements of the pair (this pseudodistance will initially be equal to the true distance between the endpoints but will get smaller each time the endpoint pair recurs).
    \item An integer $c$ which specifies the pseudodistance below which it is permissible to join pairs of endpoints. 
    \item An integer $r$ which specifies the amount by which we should reduce the pseudodistance between a pair of endpoints if we do not choose to join them. 
\end{itemize}
\vspace{-5pt}

\textbf{Subroutines:} 
\vspace{-5pt}
\begin{itemize}
    \setlength\itemsep{-0.5em}
    \item A standard MWPM decoder \textsf{MWPM} which is allowed to match endpoints to each other and to the side $X$ boundaries but not the top or bottom $X$ boundaries. It takes a vector of endpoint locations as an argument and returns a vector of endpoint location pairs (or endpoint locations and boundaries). 
    \item A function $d$ which takes as argument an endpoint location pair (or an endpoint location and a boundary) and returns the (true) distance between them.
    \item A function $t$ which takes as argument an endpoint location pair (or an endpoint location and a boundary) and returns the corresponding pair of locations in the slice for the next timestep. 
\end{itemize}
\vspace{-5pt}

\begin{algorithm}
    \caption{\textsc{Delayed Matching Decoder}}
    \label{alg:dmd}
    \begin{algorithmic}[1]
        \Require $e$, $M$
        \Ensure None
        \State $\textit{currentPairs} \gets \textsf{MWPM}(e)$
        \For{$\textit{pair} \in M$} \Comment{$M$ update step 1}
            \If{$\textit{pair} \notin \textit{currentPairs}$}
                \State Remove \textit{pair} from $M$
            \EndIf
        \EndFor
        \For{$\textit{pair} \in \textit{currentPairs}$} \Comment{$M$ update step 2}
            \If{$\textit{pair} \notin M$}
                \State $M[\textit{pair}] \gets d(\textit{pair})$
            \EndIf
        \EndFor
        \For{$\textit{pair} \in M$} \Comment{Syndrome repair step}
            \If{$M[\textit{pair}] \leq c$}
                \State Join the elements of \textit{pair} to each other
            \Else
                \State Join all endpoints in \textit{pair} to the top boundary
                \State $M[t(\textit{pair})] \gets M[\textit{pair}] - r$
            \EndIf
            \State Remove \textit{pair} from $M$
        \EndFor
    \end{algorithmic}
\end{algorithm}

Once this algorithm is complete we should have a valid $Z$ stabiliser syndrome consisting entirely of unbroken loops. We then need to use this syndrome to infer a correction on the data qubits which will push these loops to the top boundary as in Fig.~\ref{fig:transfer}. We find that the sweep decoder~\cite{kubica_cellular-automaton_2019,vasmer_cellular_2021} is a natural fit for this problem as its action is to generate corrections which push syndrome loops in a given direction, but in principle any valid decoder for loop-like syndromes could be used (e.g.~\cite{breuckmann2017,breuckmann2018,duivenvoorden2019,aloshious2019,panteleev_degenerate_2021,roffe_decoding_2020,quintavalle_single-shot_2021}). 

\begin{figure}
    \begin{subfigure}{.5\textwidth}
        \includegraphics[width=\textwidth]{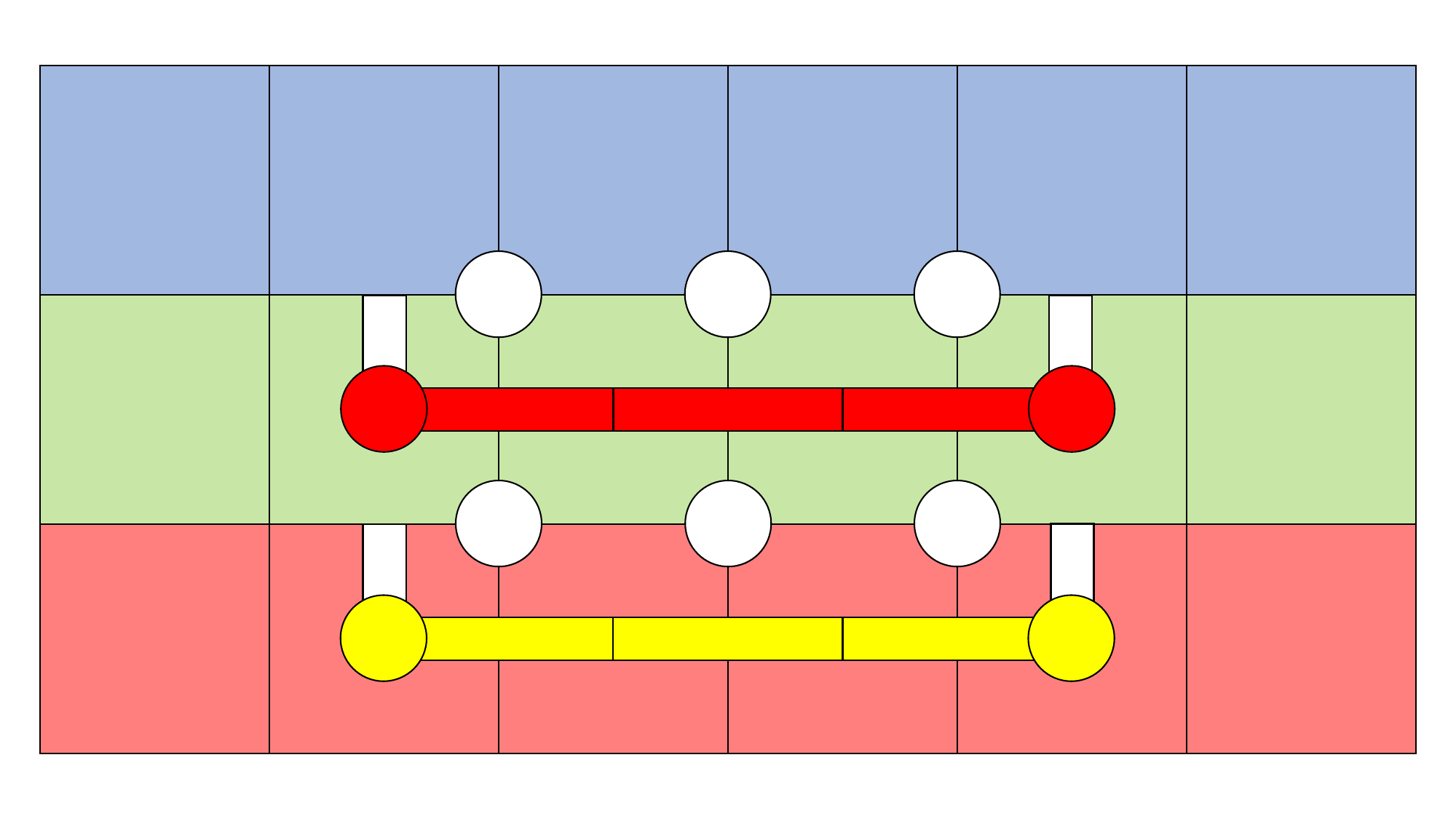}
        \subcaption{}
    \end{subfigure}
    \begin{subfigure}{.5\textwidth}
        \includegraphics[width=\textwidth]{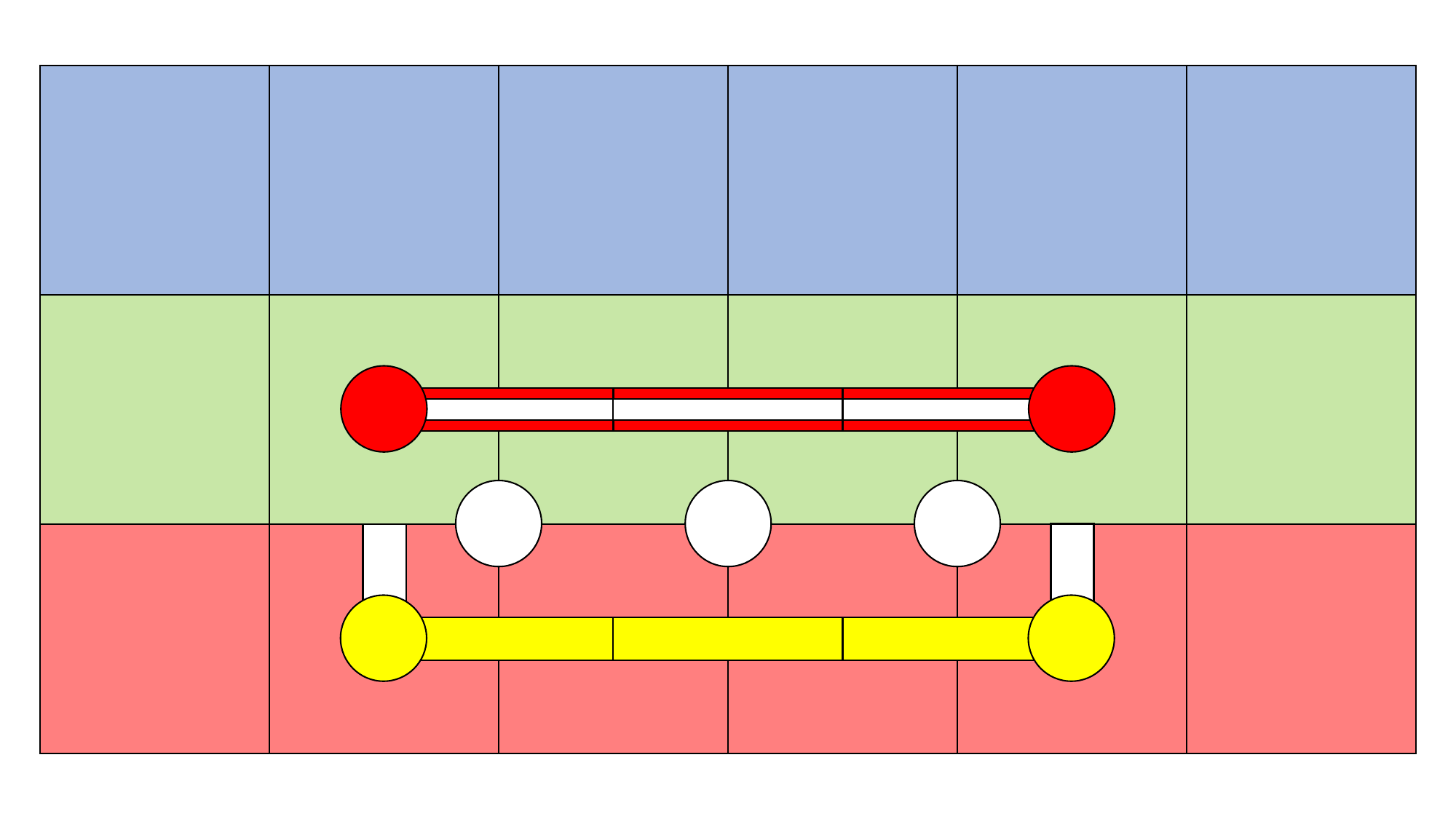}
        \subcaption{}
    \end{subfigure}
    \caption{1+1 dimensional cross-sections through (idealised) slices of a 3D surface code. Qubits are on vertices, $Z$ stabilisers are on edges (with syndromes on dual edges) and endpoints are on faces. The past, present and future are shown in red, green and blue respectively. (a) Issues caused by attempting to use a standard decoder in a sliced surface code. Three measurement errors (yellow lines) in the past caused two endpoints (yellow circles) and the decoder matched these to the top layer (white lines). An $X$ ``correction'' was then erroneously applied to three data qubits (white circles) due to this incorrect matching of endpoints. In the present, these $X$ errors cause the red syndrome and because bottom-layer stabilisers are not remeasured we get the two red endpoints. These will be matched to the top layer as before and the cycle will repeat, causing an error of unbounded size extending into the future. (b) The action of the delayed matching decoder on the same error. Once again endpoints in the past are matched to the top face, but in the present the decoder identifies that the same two endpoints occurred in the previous step and chooses to match them to each other instead of the the top.}
    \label{fig:error}
\end{figure}

It may be helpful at this point to consider the simple example presented in Fig.~\ref{fig:error}. In this case we imagine that $c = r = 2$. At the start of the procedure (in the red slice) $e$ contains the two yellow endpoints and $M$ is empty. 
\textsf{MWPM} will pair the two endpoints to each other since it is not permitted to match them to the top or bottom boundaries and the side boundaries are too far away to be favourable.
In the first update step nothing happens and in the second update step the pair of endpoints are added to $M$ with an associated value of 3 (the distance between them on dual edges). In the syndrome repair step we compare $c$ to the cost of joining the pair to each other ($M[\textit{pair}] = 3$) and since the former is smaller we join the pair to the top. A new pair of endpoints $t(\textit{pair})$ which corresponds to the yellow pair translated into the next slice (i.e. the two red endpoints) is added to $M$ with an associated value of $M[\textit{pair}] - r = 3 - 2 = 1$ and \textit{pair} is removed from $M$.

At the start of the next step (in the green slice) $e$ contains the two red endpoints and $M$ contains this pair of endpoints with an associated pseudodistance of 1. 
\textsf{MWPM} will once again match this pair to each other and nothing happens in either update step because the pair of endpoints in $M$ exactly matches the output of \textsf{MWPM}. 
In the syndrome repair step we now have $M[\textit{pair}] = 1$ which is less than $c$ so we join the two red endpoints to each other. 

\subsection{Numerical Implementation}

We numerically investigate the performance of the delayed matching decoder for all three surface codes using the slices described in Section~\ref{section:codes} and a phenomenological noise model.
Our code uses the Blossom V implementation of MWPM due to Kolmogorov~\cite{Kolmogorov2009}.
The numerical results presented here correspond to the case where we perform the expand $\rightarrow$ decode $\rightarrow$ collapse parts of the procedure for $O(d)$ cycles but do not apply $CCZ$s between the codes. 
As such, we include both measurement errors on $Z$ stabilisers and physical $X$ errors in our simulation but not $Z$ errors or $CZ$ errors, which arise due to application of $CCZ$ to qubits with $X$ errors. We use parameters $c=2$ and $r=1$ for the delayed matching decoder. The code used to obtain these results can be found at \cite{gitrepo_code}.

For each of the three codes, we perform Monte Carlo simulations of the procedure for different values of $p$, the $X$ error and measurement error rate, and $L$, the code distance. For each combination of parameters, we count the number of failure events $f$ that occur in a number of trials $n$, and estimate the logical error rate $p_{\rm fail}(p, L)$. Rather than using the standard confidence interval based on the normal distribution, we use the more stable Agresti-Coull confidence interval~\cite{Agresti1998}. Let $\tilde f = f + 2$, $\tilde n = n + 4$, and $\tilde p = \tilde f / \tilde n$.  
Our estimate of $p_{\rm fail}(p, L)$ is then:
\begin{equation}
    p_{\rm fail}(p, L) = \tilde p \pm 2 \sqrt{\frac{\tilde p (1 - \tilde p)}{\tilde n}},
\end{equation}
which is approximately the 95\% confidence interval. 

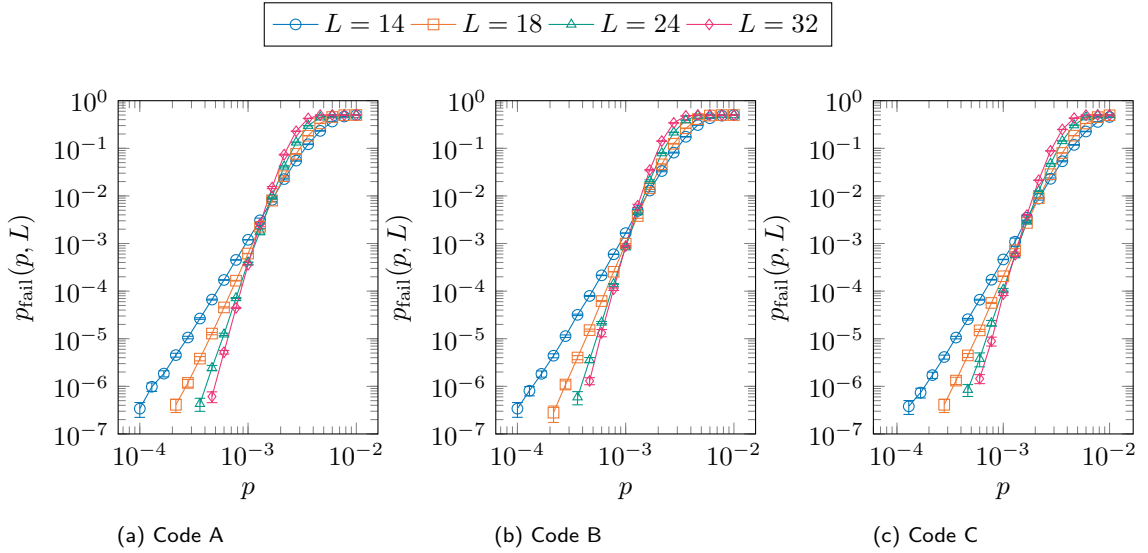
\begin{figure}
    \centering
    \begin{tikzpicture} 
        \begin{axis}[%
        hide axis,
        xmin=10,
        xmax=50,
        ymin=0,
        ymax=0.4,
        legend columns=4
        ]
        \addlegendimage{cb-blue,mark=o}
        \addlegendentry{$L=14$};
        \addlegendimage{cb-orange,mark=square}
        \addlegendentry{$L=18$};
        \addlegendimage{cb-teal,mark=triangle}
        \addlegendentry{$L=24$};
        \addlegendimage{cb-magenta,mark=diamond}
        \addlegendentry{$L=32$};
        \end{axis}
    \end{tikzpicture}
    \par\bigskip
    \begin{subfigure}{.3\textwidth}
    \centering
        \begin{tikzpicture}
            \begin{axis}[
                xlabel={$p$},
                ylabel={$p_{\mathrm{fail}}(p,L)$},
                ymode=log,
                xmode=log,
                legend style={
                    at={(0.95,0.05)},
                    anchor=south east,
                    nodes={scale=0.9, transform shape}
                },
                ytick={
                    0.0000001,0.000001,0.00001,0.0001,0.001,0.01,0.1,1
                },
                width=5cm,
                height=6cm,
                ymin=1e-7,
                ymax=1
            ]
            \addplot[
                color=cb-blue,
                mark=o,
                mark size=2,
                error bars/.cd, 
                y dir=both, 
                y explicit
            ] 
            table [x=p, y=pfail, y error=err95, col sep=comma] {data/L14_Toric.csv};
            \addplot[
                color=cb-orange,
                mark=square,
                mark size=2,
                error bars/.cd, 
                y dir=both, 
                y explicit
            ] 
            table [x=p, y=pfail, y error=err95, col sep=comma] {data/L18_Toric.csv};
            \addplot[
                color=cb-teal,
                mark=triangle,
                mark size=2,
                error bars/.cd, 
                y dir=both, 
                y explicit
            ] 
            table [x=p, y=pfail, y error=err95, col sep=comma] {data/L24_Toric.csv};
            \addplot[
                color=cb-magenta,
                mark=diamond,
                mark size=2,
                error bars/.cd, 
                y dir=both, 
                y explicit
            ] 
            table [x=p, y=pfail, y error=err95, col sep=comma] {data/L32_Toric.csv};
            \end{axis}
        \end{tikzpicture}
        \subcaption{Code A}
    \end{subfigure}
    ~~
    \begin{subfigure}{.3\textwidth}
    \centering
        \begin{tikzpicture}
            \begin{axis}[
                xlabel={$p$},
                ylabel={$p_{\mathrm{fail}}(p,L)$},
                ymode=log,
                xmode=log,
                legend style={
                    at={(0.95,0.05)},
                    anchor=south east,
                    nodes={scale=0.9, transform shape}
                },
                ytick={
                    0.0000001,0.000001,0.00001,0.0001,0.001,0.01,0.1,1
                },
                width=5cm,
                height=6cm,
                ymin=1e-7,
                ymax=1
            ]
            \addplot[
                color=cb-blue,
                mark=o,
                mark size=2,
                error bars/.cd, 
                y dir=both, 
                y explicit
            ] 
            table [x=p, y=pfail, y error=err95, col sep=comma] {data/L14_Rhombic1.csv};
            \addplot[
                color=cb-orange,
                mark=square,
                mark size=2,
                error bars/.cd, 
                y dir=both, 
                y explicit
            ] 
            table [x=p, y=pfail, y error=err95, col sep=comma] {data/L18_Rhombic1.csv};
            \addplot[
                color=cb-teal,
                mark=triangle,
                mark size=2,
                error bars/.cd, 
                y dir=both, 
                y explicit
            ] 
            table [x=p, y=pfail, y error=err95, col sep=comma] {data/L24_Rhombic1.csv};
            \addplot[
                color=cb-magenta,
                mark=diamond,
                mark size=2,
                error bars/.cd, 
                y dir=both, 
                y explicit
            ] 
            table [x=p, y=pfail, y error=err95, col sep=comma] {data/L32_Rhombic1.csv};
            \end{axis}
        \end{tikzpicture}
        \subcaption{Code B}
    \end{subfigure}
    ~~
    \begin{subfigure}{.3\textwidth}
    \centering
        \begin{tikzpicture}
            \begin{axis}[
                xlabel={$p$},
                ylabel={$p_{\mathrm{fail}}(p,L)$},
                ymode=log,
                xmode=log,
                legend style={
                    at={(0.95,0.05)},
                    anchor=south east,
                    nodes={scale=0.9, transform shape}
                },
                ytick={
                    0.0000001,0.000001,0.00001,0.0001,0.001,0.01,0.1,1
                },
                width=5cm,
                height=6cm,
                ymin=1e-7,
                ymax=1,
                xmin=6e-5
            ]
            \addplot[
                color=cb-blue,
                mark=o,
                mark size=2,
                error bars/.cd, 
                y dir=both, 
                y explicit
            ] 
            table [x=p, y=pfail, y error=err95, col sep=comma] {data/L14_Rhombic2.csv};
            \addplot[
                color=cb-orange,
                mark=square,
                mark size=2,
                error bars/.cd, 
                y dir=both, 
                y explicit
            ] 
            table [x=p, y=pfail, y error=err95, col sep=comma] {data/L18_Rhombic2.csv};
            \addplot[
                color=cb-teal,
                mark=triangle,
                mark size=2,
                error bars/.cd, 
                y dir=both, 
                y explicit
            ] 
            table [x=p, y=pfail, y error=err95, col sep=comma] {data/L24_Rhombic2.csv};
            \addplot[
                color=cb-magenta,
                mark=diamond,
                mark size=2,
                error bars/.cd, 
                y dir=both, 
                y explicit
            ] 
            table [x=p, y=pfail, y error=err95, col sep=comma] {data/L32_Rhombic2.csv};
            \end{axis}
        \end{tikzpicture}
        \subcaption{Code C}
    \end{subfigure}
    \caption{
    Numerical evidence of an error threshold for the delayed matching decoder. We ran Monte Carlo simulations for the three codes described in Section~\ref{section:codes}. For each code, we estimate the logical error rate $p_{\rm fail}(p, L)$ as a function of the $X$ error and measurement error rate, $p$, for different code distances $L$. The error threshold, $p_c$, is the value of $p$ where the curves for different $L$ intersect. We observe an error threshold of $p_c \sim 0.1\%$ in each of the three codes.
    }
    \label{fig:threshold_plots}
\end{figure}

For each code, we plot $p_{\rm fail}$ against $p$ for different values of $L$. The error threshold, $p_c$, is the point where the curves for different $L$ intersect. We observe evidence of an error threshold in the region of $p_c \sim 0.1 \%$ for all three codes; see Fig.~\ref{fig:threshold_plots}. 
Our error threshold estimate is many orders of magnitude larger than the theoretical value of $\sim 10^{-17}$ obtained by Brown in his proof~\cite{brown_fault-tolerant_2020}. 
In addition, our value is only one order of magnitude smaller than the 2D surface code threshold~\cite{raussendorf2007,wang2011,fowler_surface_2012,stephens2014}.
We anticipate that the performance of the delayed matching decoder could be significantly improved by using more of the syndrome history, so it is still possible that the optimised JIT decoding threshold could be competitive with the 2D surface code threshold.
We observe improved suppression of $p_{\rm fail}$ below threshold in code C when compared to codes A and B. This is consistent with the fact that, for a given value of $L$, $\overline{X}$ is higher weight in the layers for this code than in the other two. 

We also analyse the behaviour of the logical error rate below the threshold.
We assume the following ansatz for the logical error rate for $p < p_c$,
\begin{equation}
    p_{\mathrm{fail}}(p, L) \propto p^{\alpha L^\beta}.
    \label{eq:subthreshold}
\end{equation}
We use the fitting method detailed in~\cite{brown2016b} to estimate $\alpha$ and $\beta$ for all three codes; see Table~\ref{tab:subthreshold} for the results. 
Our estimates are consistent with the statement that (to leading order) logical failures are caused by errors of weight $O(d)$ where $d=L$ is the code distance. 

\begin{table}[h]
    \centering
    \begin{tabular}{c|ccc}
        & Code A & Code B & Code C \\
        \hline
        $\alpha$ & $0.29(4)$ & $0.23(8)$ & $0.21(5)$ \\
        $\beta$ & $0.97(4)$ & $1.08(11)$ & $1.09(8)$ \\
    \end{tabular}
    \caption{Estimated parameters for the sub-threshold scaling of $p_{\mathrm{fail}}$, using the ansatz in Eq.~\eqref{eq:subthreshold}.}
    \label{tab:subthreshold}
\end{table}

\section{Discussion}
\label{section:discussion}
Now that we have covered each component of Brown's procedure in detail, it is helpful to once again provide an overview of the procedure so that we can see how each component fits together. A single timestep proceeds as follows:

\begin{enumerate}
    \item Begin with three 2D surface codes. All $Z$ stabilisers of these codes are assumed to be in the +1 eigenstate.
    \item Expand to three thin slices of 3D surface code by preparing new data qubits in $\ket{+}$ and then measuring the new $Z$ stabilisers. The previously existing 2D codes will now be the bottom layers of these slices, and their stabilisers are \textit{not} remeasured. 
    \item If our assumption regarding the states of the initial $Z$ stabilisers was correct and there are no measurement errors on the newly measured ones then the syndrome from these stabilisers will consist entirely of loops in the bulk or connected to the top or side boundaries. A correction that pushes all these loops to the top boundary will transfer the original state of the 2D code into the slice. If there are measurement errors on the newly measured stabilisers or X errors on the lower layer we will have an invalid syndrome containing broken strings, the endpoints of which must be paired up to produce a valid syndrome from which we can infer a correction. Incorrect pairings will result in X errors in the slice. This is the JIT decoding step.
    \item Apply CCZs between all triples of qubits in the region where the three slices overlap except for those in the top layer. If there are X errors in this region due to incorrect decoding in the previous step then these will cause CZ errors on the other two codes. 
    \item Measure out all non-top layer qubits in the X basis. The output of these measurements will be fed to a global Z error decoder once the procedure is complete and used to find a Z error correction for the final code.
\end{enumerate}

We have verified the integrity of the majority of the components involved in this process. The only part which is missing is the simulation of the Clifford errors arising from the non-Clifford gate and the numerical demonstration of a threshold for $Z$ errors. Such simulations will be the topic of future work, and will allow us to compare the performance of the JIT non-Clifford gate against the standard magic state distillation approach~\cite{bravyi2005,fowler_surface_2012,Litinski2019magicstate}. 
We wish to emphasize that while the Clifford errors may affect the overall threshold for the procedure, they will not affect the JIT decoding thresholds shown in this work.
This is because the JIT decoder only deals with $X$ errors and the errors arising from the $CCZ$ gate are $CZ$ errors which will be projected to a distribution of $Z$ errors by the single-qubit measurements which collapse the slice. 

In contrast, the locations of the $CZ$ errors depend on the locations of $X$ errors post-JIT decoding, so while the distribution of $Z$ and $CZ$ errors will not affect the performance of the JIT decoder the reverse is not true. For this reason the development of more sophisticated and effective JIT decoders is also an important direction for future research. The delayed matching decoder has the advantage of being relatively simple, but it uses only a small amount of the information available in the syndrome history and it is reasonable to expect that significant improvements could be made to decoder performance by utilising more of the available information. 

Another natural direction for future research would be the construction of similar slices in other topological codes; for example, one might want to perform an analogous procedure in the 2D/3D colour code in order to perform a linear-time logical T gate. However, this is not quite as straightforward as one might hope because, unlike in the 3D surface code,  the string-like logical operator of the 3D tetrahedral colour code is not required to run in any particular direction and any edge of the tetrahedron supports a valid implementation of this operator. This means that any slice of bounded height through the 3D code will contain edges of bounded length which support low-weight logical operator implementations. Cubic colour code constructions encoding multiple logical qubits as in~\cite{kubica_unfolding_2015} can avoid this problem for some but not all logical qubits. In these codes, as with three surface codes admitting a transversal $CCZ$, the string-like logical operators are all perpendicular, but unlike in the surface code case we cannot assign different time directions to different logical qubits. This means that for any choice of slice the string-like logical for one of the encoded qubits will run between the top and bottom boundaries. Additionally, this logical qubit will be lost completely in the collapse from 3D to 2D as the corresponding square 2D colour code only encodes two qubits. This does not completely rule out more exotic slices in these codes (e.g. with logical qubits encoded in topological defects rather than code boundaries) but it seems unlikely that slices similar to the ones presented here exist for 3D colour codes. 

It is worth emphasising the significant difference between the JIT decoding scheme we have examined in this work and Bomb\'{i}n's original JIT decoding proposal~\cite{bombin_2d_2018}. In particular, the above arguments regarding the difficulties of constructing valid colour code slices do not apply to that scheme because it uses an measurement-based formalism rather than a circuit-based one and so does not involve dimension jumping operations. These operations are what would map low-weight logicals in the slice to weight-$d$ logicals in the 2D code post-collapse, but the measurement-based formalism allows for a continuous ``sliding'' of the slice through the 3D spacetime and so dimension jumps are not required and these short error strings will be detected as we slide the slice past them. The JIT decoder proposed by Bomb\'{i}n in~\cite{bombin_2d_2018} also differs significantly from the one discussed here.

In light of recent results regarding transversal $CCCZ$ gates in the 4D surface code~\cite{kubica_unfolding_2015,jochym-oconnor_four-dimensional_2021} it also seems natural to ask if an equivalent process could be used to construct slices of 4D surface code which allow for a linear-time $CCCZ$ in the 3D surface code. We expect that such a generalisation should be possible, but note that its spacetime overhead would scale as $d^4$. This compares unfavourably with the $d^3$ scaling allowed by the constant-time set of computationally universal operations in the 3D surface code~\cite{vasmer_three-dimensional_2019}, and so is unlikely to be advantageous.

\section*{Acknowledgements}

TRS acknowledges support from University College London and the Engineering and Physical Sciences Research Council [grant number EP/L015242/1]. The authors acknowledge the use of the UCL Myriad High Performance Computing Facility (Myriad@UCL), and associated support services, in the completion of this work. Research at Perimeter Institute is supported in part by the Government of Canada through the Department of Innovation, Science and Economic Development Canada and by the Province of Ontario through the Ministry of Colleges and Universities. This research was enabled in part by support provided by \href{www.computeontario.ca}{Compute Ontario} and \href{www.computecanada.ca}{Compute Canada}. PW acknowledges support from the Australian Research Council via the Centre of Excellence in Engineered Quantum Systems (EQUS) project number CE170100009. DEB acknowledges support from the Engineering and Physical Sciences Research Council QCDA project [grant number EP/R043647/1].

The authors thank Ben Brown for illuminating discussions. TRS thanks T. Davies for assistance with 3D models and presentation of figures.

\bibliographystyle{unsrtnat}
\bibliography{references}

\end{document}